\documentclass{ametsocV6.1}

\usepackage{booktabs, tabularx}
\usepackage{multirow}

\title{Dry-to-Wet Soil Gradients Enhance Convection and Rainfall over Subtropical South America}

\authors{Divyansh Chug\thanks{Divyansh Chug's current affiliation: Center for Policy Research on Energy and the Environment, Princeton University},\aff{a}\correspondingauthor{Divyansh Chug, dchug2@illinois.edu} 
Francina Dominguez,\aff{a} 
Christopher M. Taylor,\aff{b,c} 
Cornelia Klein,\aff{b,d} 
Stephen W. Nesbitt\aff{a} 
}

\affiliation{\aff{a}{Department of Atmospheric Sciences, University of Illinois Urbana Champaign, Urbana, Illinois}\\
\aff{b}{U.K. Centre for Ecology and Hydrology, Wallingford, United Kingdom}\\
\aff{c}{National Centre for Earth Observation, Wallingford, United Kingdom}\\
\aff{d}{Institute of Atmospheric and Cryospheric Sciences, University of Innsbruck, Innsbruck, Austria}
}

\abstract{Soil moisture-precipitation (SM-PPT) feedbacks at the mesoscale represent a major challenge for numerical weather prediction, especially for subtropical regions that exhibit large variability in surface SM. How does surface heterogeneity, specifically mesoscale gradients in SM and land surface temperature (LST), affect convective initiation (CI) over South America? Using satellite data, we track nascent, daytime convective clouds and quantify the underlying antecedent (morning) surface heterogeneity. We find that convection initiates preferentially on the dry side of strong SM/LST boundaries with spatial scales of tens of kilometers. The strongest alongwind gradients in LST anomalies at 30 km length scale underlying the CI location occur during weak background low-level wind ($<$2.5m/s), high convective available potential energy ($>$1500J/kg) and low convective inhibition ($<$250J/kg) over sparse vegetation. At 100 km scale, strong gradients occur at the CI location during convectively unfavorable conditions and strong background flow. 
The location of PPT is strongly sensitive to the strength of the background flow. The wind profile during weak background flow inhibits propagation of convection away from the dry regions leading to negative SM-PPT feedback whereas strong background flow is related to longer lifecycle and rainfall hundreds of kilometers away from the CI location. Thus, the sign of the SM-PPT feedback is dependent on the background flow.
This work presents the first observational evidence that CI over subtropical South America is associated with dry soil patches on the order of tens of kilometers. Convection-permitting numerical weather prediction models need to be examined for accurately capturing the effect of SM heterogeneity in initiating convection over such semi-arid regions.} 

\begin{document}

\maketitle

\section{Introduction} \label{intro}

The boundary layer evolution over a heterogeneously moist surface is different from that over uniform terrain due to the spatial variability of surface buoyancy flux and the resultant wind circulation \citep{Pielke1991, segel1992, taylor_observational_2007, lee_effect_2019}. This is of particular importance for ``transition zones" between dry and wet climatic regimes where evaporation is limited by soil moisture \citep{koster2004}, which controls the partitioning of surface energy into heat and moisture fluxes. Contingent on the background wind and the length scale of surface heterogeneity, the induced circulation can increase the spatial variation in boundary layer depth, mixed layer potential temperature and convective forcing \citep{ChenAvissar1994, ChengCotton2004}, and provide vertical motion triggers that can initiate deep convection \citep{taylor_frequency_2011}. The role of soil moisture (SM) heterogeneity and mesoscale circulations in convective initiation (CI) has not been studied over subtropical South America. 

Numerical modeling studies using cloud-resolving models or large-eddy simulations have shown that convergence and moist ascent is maximized over strong dry-wet SM boundaries of an idealized heterogeneous surface and convective triggering preferentially occurs near the downwind end of the drier surface region \citep{SegalArritt1992, ChenAvissar1994, ChengCotton2004, Adler_inhomogeneities_2011, Garcia_InducedFlows_2011, KangBryan2011, Rieck_Clouds_2014, Rochetin_Breeze_2017, lee_effect_2019}. The strength of this induced flow is sensitive to the length scale of SM heterogeneity, the magnitude of temperature gradient at the surface and the strength of background wind. Using computationally-limited domain sizes ($<$200 km) and checkerboard-style representation of surface heterogeneity, these studies have converged on a numerically-determined empirical limit of 3-5 m/s for the strongest background wind. This framework suffers major drawbacks. The checkerboard representation of surface heterogeneity does not provide a true picture of how this process occurs in nature. The simulated maximum patch sizes of up to 40km is limited by domain size. We hypothesize that the determined limit of background wind speed is a byproduct of the patch size limitation. In the real world, SM heterogeneity can occur over a spectrum of length scales from tens to hundreds of kilometers. The observed spectrum of length scales of SM heterogeneity associated with CI has not been studied in the context of mesoscale circulations.

Following CI over drier soils, the soil moisture-precipitation (SM-PPT) feedback can vary in strength as well as sign due to the background conditions and the length scale considered \citep{emori_feedback_1998, schar_feedback_1999}. For example, \cite{froidevaux_wind_feedback_2014} used an idealized cloud-resolving framework to show positive feedback when convection initiates over dry soil patches but the convective cells with longer lifetime propagate over wet soil patches, strengthen and preferentially precipitate. Negative feedback occurs when a weak wind profile cannot support the propagation of convective features from dry to wet areas leading to more rainfall over the dry patch itself. Understanding the role of SM anomalies in afternoon convection and accurately quantifying convective scale SM-PPT feedbacks is critical for the prediction of PPT from diurnal to seasonal time-scales as convective parameterizations and spatial scale may confound both the feedback strength and the sign \citep{tuttle_feedback_2016}. This scale-dependence of SM-PPT feedbacks has been demonstrated in case of numerical model output but observational support is lacking. 

Recent work has used remotely-sensed observations to extract the ``fingerprint" of SM heterogeneity in CI occurrence and PPT accumulation. Variations in SM at length scales of 10-40 km exert a strong control on storm initiation in the Sahel \citep{taylor_frequency_2011}, Europe \citep{taylor_detecting_2015} and the Tibetan plateau \citep{barton_observed_2021}. \cite{taylor_frequency_2011} showed that 37\% of all analyzed storm initiations over the Sahel region occurred over the steepest 25\% of SM gradients. Global analyses of mesoscale SM patterns antecedent to afternoon PPT show a clear preference for rainfall over spatial minima in SM anomalies \citep{taylor_afternoon_2012, guillod_reconciling_2015}.  These studies provide strong evidence for preferential daytime convection and rainfall over spatially drier patches of soil of the order of tens of kilometers in semi-arid regions.


Daytime convection is the most important mechanism contributing to seasonal rainfall accumulation over subtropical South America, specifically over the La Plata river basin (LPB). Organized convective storms can contribute more than 70\% of the rainfall in this region \citep{nesbitt_trmm_2006, rasmussen_contribution_2016}. These storms typically originate in early afternoon hours that coincides with the theoretical maximum probability of CI due to SM heterogeneity-induced mesoscale circulations. Maximum PPT accumulation is observed in late night and early morning hours when these storms progress eastward and can last up to 18-24 hours after initiation \citep{rasmussen_contribution_2016}. Past studies have highlighted the importance of SM in driving the spatiotemporal variation of ET, air temperature and PPT \citep{Baker_LA_2021} over South America. Interactions between the land surface and the atmosphere are important drivers of weather and climate over subtropical South America, one of the ``hotspots" of land-atmosphere coupling \citep{koster2004, sorensson_summer_2011, ruscica_pathways_2015}. Typical coarse-grid analyses cannot distinguish the effect of land-atmosphere interactions during the different stages of convection (from initiation to dissipation) and yield an incomplete picture of the role of surface feedbacks in regional PPT. The aforementioned ``fingerprint" technique applied to high-resolution observations over South America can fill the identified gaps in convective-scale SM feedback studies.

The goals of this study are: a) quantify the relationship between daytime convection and the antecedent, underlying, SM anomalies, b) formulate data-driven hypothesis on how background geographical and atmospheric conditions affect the SM-CI relationship, and c) analyze the controls on the strength and sign of SM-PPT feedback for CI over drier soils. We use satellite observations of infrared brightness temperature (T${_b}$), surface SM, land surface temperature (LST) and PPT to examine the influence of mesoscale SM variability on the initiation of convection and subsequent rainfall during the warm season (November-March; NDJFM) in South America. 

\begin{figure}[!h]
\centering
\vspace{\baselineskip}
\includegraphics[width=\textwidth]{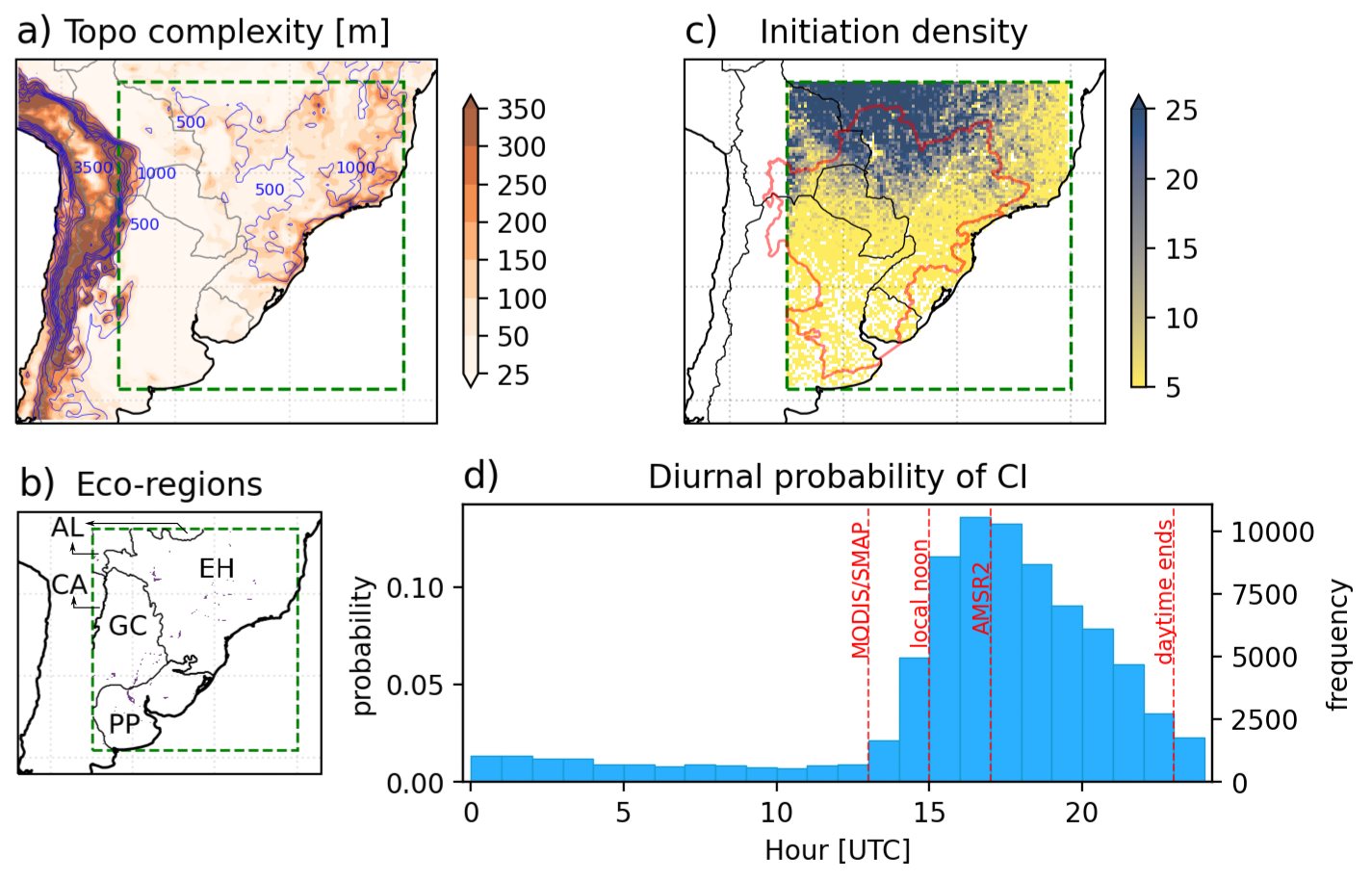}
\caption{\label{domain}a) Elevation (m; blue contours) and standard deviation of elevation (m; orange shading) over every 40km x 40km area (i.e. topographic complexity). Green dashed box shows the domain for the analysis. b) Eco-regions within the domain denoted by AL (Amazonian Lowlands), CA (Central Andes), EH (Eastern Highlands), GC (Gran Chaco) and PP (Pampas Pantanal) and location of lake bodies greater than 10 km$^{2}$ in area (shading). c) Frequency of detected CI events resampled to a 0.25$^{\circ}$ x 0.25$^{\circ}$ grid. Red delineated boundary shows the La Plata basin. c) Diurnal probability distribution of tracked CI events; the red dashed lines from left to right show the start time of CI events considered for MODIS and SMAP analysis, local noon, the start time of CI events considered from AMSR2 analysis and the end of daytime respectively.}
\end{figure}

\section{Data and Methods} \label{methods}

\subsection{Study region}

The La Plata basin (LPB) in South America covers 3.17 million km$^{2}$ hydrographical area between 14$^{\circ}$S-38$^{\circ}$S latitudes (red delineated boundary in Fig. \ref{domain}c). The land surface over LPB is dominantly covered by savannas, grasslands and croplands along with deciduous broadleaf forests in southeastern Bolivia and southeastern Brazil. Both PPT and vegetation density generally decrease from north to south and from east to west within the basin with a maximum located in central-east LPB at the climatological location of the convergence of the northerly moisture flux transported by the South American low-level jet \citep[see Fig. 1 of][]{chug_isolating_2019}. The domain for this analysis broadly encompasses LPB (green dashed box in Fig. \ref{domain}a). The elevation within the domain ranges up to 3200 $m$ and the topographic complexity (TC), defined as the local standard deviation of elevation in every 40 km-by-40 km area, ranges from 2 $m$ to 1074 $m$ (orange shading in Fig. \ref{domain}a). 

\subsection{Tracking nascent-stage convective features}
To understand SM feedbacks to CI, we need to detect the locations and timings of the first initiation of deep convective clouds. Convective clouds can be detected from infrared satellite images as local minima of brightness temperature which are strongly correlated with deep convection. We use the half-hourly infrared brightness temperature data (T${_b}$; equivalent blackbody temperature) at 4 km horizontal resolution globally-merged from multiple geostationary satellites \cite{Janowiak_GOES_2017}. This dataset has been pre-corrected for zenith angle dependence to reduce the cold bias at locations far from satellite nadir.

Following \cite{barton_observed_2021}, we define a procedure to detect deep convective clouds in early stages of growth such that the horizontal area of the detected clouds is always smaller than or equal to a maximum size threshold, chosen to be 20 km-by-20 km in this study. This limits the cloud window to a 5 x 5 grid in the satellite T${_b}$ field at 4 km resolution. The procedure tracks the location of such cloud windows in the T${_b}$ field that have a cold core, i.e., all 5 x 5 grids where the coldest temperature is found at the central pixel. This coldest temperature must be less than a threshold core temperature for deep convective clouds (235 K in this study). All such detected windows represent potential convective clouds in early stages of growth.

To reduce errors due to pre-existing and moving clouds, the detected clouds need to be newly emergent and rapidly cooling. Hence, we apply the following additional criteria: 
\begin{enumerate}
  \item To ensure a newly emergent cloud, the periphery of the 5 x 5 feature in the time step of detection should be free of cloud-related pixels. Each pixel within 12 km of the cloud window should be warmer than a maximum brightness temperature found in a cloud layer. This highest cloud layer T${_b}$ associated with deep convection is set to 253K \citep{MapesHouze1993}.
  \item To ensure that this feature is rapidly cooling and not associated with a moving cloud, we check the progression of minimum T${_b}$ within the cloud window and its periphery. We find the linear regression slope (K hr$^{-1}$) of the minimum T${_b}$ over the last 1 hour using 2 earlier time steps. This slope must be less (more negative) than a threshold slope that represents rapid cooling (-10K hr$^{-1}$ in this study). 
\end{enumerate} 
  
If all criteria are satisfied, the center of the detected cloud feature is recorded as the location of CI along with the respective timing of detection. This algorithm is applied to all 5 x 5 windows in the domain during the local daytime hours (13 UTC-23 UTC or 10 am-8pm local time) for the wet season over subtropical South America (November through March; NDJFM) over the period 2001-2020. Parallax correction is applied to reduce the error in sensing the ground location of tracked cloud features. We estimate the approximate cloud-top heights using monthly mean temperature profile and geopotential heights from ERA5 reanalysis \citep{hersbach2020}. To minimize the impact of convection over or related to water bodies, all CI events within 10 $km$ of a water body greater than 10 $km^{2}$ in area (Fig. \ref{domain}b) are removed. This leads to a 10\% reduction in detected convective events within the domain.

\subsection{Composite patterns of antecedent surface wetness}

After detecting daytime CI events over the domain, the next step is to quantify the antecedent surface anomalies underlying each event. There is no single measurement system that can accurately map SM and the associated surface flux variability at a resolution appropriate to study convective feedbacks. Hence, we employ three different satellite datasets. 

\subsubsection{Surface soil moisture from microwave radiometer}

Daily morning surface SM (units: cm$^{3}$cm$^{-3}$) estimates at approximately 6 am local time from the Soil Moisture Active Passive (SMAP) enhanced level-3 radiometer dataset \citep{SMAP_L3_v5} at 9 km horizontal resolution are employed to extract SM anomaly (SMA) patterns underlying the detected CI events. This dataset is available from 2015 onward. The SMAP SM estimates are derived from the natural thermal emission of the Earth's surface at low microwave frequency ($\sim$1 GHz), providing the following advantages: a) This frequency lies within the spectral window where the atmosphere is nearly transparent to outgoing energy, b) measurements are independent of solar illumination, and c) thermal emission from the soil underlying vegetation can be retrieved for sparse to moderate foliage, up to atleast 5 kg m$^{-2}$ vegetation water content \citep{SMAP_L3_v5}. 

A 31-day running mean climatology over all years (2015-2020) centered at each calendar day is subtracted from the respective day to calculate SMA. We adopt a similar methodology to \cite{taylor_frequency_2011} and \cite{barton_observed_2021}. SMA over a surface transect of 250 km-by-250 km centered at the closest SMAP grid point directly underlying the CI location was extracted for each CI event. The analysis is limited to cases with the highest quality data at the CI location.

\subsubsection{Land surface temperature from thermal infrared radiometer}
Daily morning observations of LST (units: K) at approximately 10:30 am local time from level-3 MODIS/Terra dataset version 6.1 at 0.05$^{\circ}$ ($\sim$5600 m at equator) horizontal resolution \citep{MOD11C1_061} are used for the entire duration of the study (2001-2020). MODIS LST is available for longer duration at a higher spatial resolution compared to SMAP SM, provides more detailed retrieval quality description and is directly linked to the surface energy budget. Infrared LSTA is employed as a proxy for spatial variability in SM because moister soils are associated with a cooler temperature due to higher specific heat capacity and exhibit higher thermal emissivity due to enhanced absorption and emission in the thermal infrared range. LST is determined from the thermal infrared signature received by the MODIS sensors in the spectral bands between 3.66-12.27 $\mu$m range after accounting for atmospheric emission, absorption and scattering of thermal radiation from the surface and from the sun. Since clouds are not transparent in this wavelength range, MODIS LST data is produced only for clear sky conditions.
We follow a similar procedure as with the SMAP dataset but with key differences. We subtracted a 15-day running mean climatology (instead of 31-day) while calculating the temporal LST anomalies (LSTAs) to account for the larger sample size of the data. Due to lack of ground validation studies of MODIS LST data over subtropical South America and the errors related to the cloud removal algorithm. i.e., frequent cold bias due to cloud shadow effects, we limit this analysis to pixels with the best quality data - a) cloud-free conditions, b) average LST error $<$ 1 K, and c) average emissivity error $<$ 0.01. To alleviate the subsequent sampling issues due to the filtering, we only include cases when - a) there is a valid retrieval underlying the initiation point, and b) the surface domain contains the best quality retrievals for more than 50\% pixels.

\subsubsection{Land surface temperature from microwave radiometer}

We employed afternoon LST observations at approximately 1:30 pm local time from the 10 km AMSR2 downscaled LPRM level-2 product \citep{AMSR2_dataset} as a third dataset to validate the surface anomaly patterns. Similar to SMAP, AMSR2 provides all-sky observations. However, its high microwave frequency (36.5 GHz) is sensitive to hydrometeors \citep{Gao_LST_2008, Holmes_LST_2009}. Hence, we filter the dataset using GPM IMERG Level 3 High Quality precipitation based on microwave-only data \citep{IMERG_hh_2019}. Following \cite{barton_observed_2021}, all AMSR2 pixels with rainfall rates greater than 1 mm hr$^{-1}$ between 1–3 pm local time are excluded from the analysis. AMSR2 LSTAs are calculated after removing a 31-day running mean climatology for 2013-2021 at each grid point. Since the study is focused on the effect of pre-storm surface heterogeneity, we sample AMSR2 LSTAs only for CI cases that initiate after 2 pm local time.
\subsubsection{Alongwind rotation and composite analysis}

Before compositing the surface anomaly transects for the respective surface datasets, each transect was rotated in the direction of the background wind. The background wind was calculated as the 10-m wind from ERA5 dataset spatially averaged over the 1$^{\circ}$x1$^{\circ}$ area centered at the CI location and temporally averaged over 2 hours preceding the corresponding initiation time. A composite of mean surface conditions was generated as the average of rotated surface transects for all cases for a respective dataset. Since the CI events occur over a range of latitudes, topography, daytime hours, etc., we subtracted the spatial mean SMA (or LSTA) from each transect before calculating the composite pattern.

\subsection{Rainfall patterns after convective onset}

Half-hourly high quality microwave-only rainfall rate estimates from the IMERG final run dataset \citep{IMERG_hh_2019} were used for analyzing the 24-hour rainfall following the CI events. This product provides PPT estimates that are intercalibrated and merged from various passive microwave sensors, and then corrected using monthly gauge-based PPT to produce the final IMERG output of rain rate at a $\sim$10 km grid. The 24-hour PPT composites corresponding to each CI event are constructed from 250 km-by-250 km transects of PPT rate centered at the location of each CI event. The alongwind composites of PPT anomaly are derived similarly as the surface anomaly composites.

We also used this dataset to ensures that our signal is not contaminated by the effect of early morning PPT on SM conditions. All CI cases with morning (7-10 am local time) PPT within the domain exceeding 1 mm hr$^{-1}$ were removed before the composite analysis. 

\subsection{Other datasets used in the study}

\begin{itemize}

\item{GMTED2010: Global Multi-resolution Terrain Elevation Data at 30 arc-second resolution was used for inferring elevation over the study domain and calculating TC, the local standard deviation of elevation over a 40 km-by-40 km region.} 

\item{ERA5: Wind vectors at 10 m above ground and multiple pressure levels in the atmosphere, convective available potential energy (CAPE) and convective inhibition (CIN) from the ERA5 dataset \citep{hersbach2020}.} 

\item{Other MODIS datasets: The Terra MODIS land water mask (MOD44W) version 6 dataset \citep{MODIS_lakes} for the most recent year (2015) was used to filter out all CI events related to water bodies. The global map of surface water at 0.25 km spatial resolution was conservatively regridded to 4 km to preserve the integral of water body area within each pixel. 

Monthly-mean Enhanced Vegetation Index (EVI) at 0.05$^{\circ}$ resolution \citep{MODIS_EVI} was used to estimate background vegetation density for each CI event. EVI is derived from atmospherically-corrected differential reflectance in the red, near-infrared, and blue wavelength bands. The usage of the blue band minimizes canopy-soil variations and improves saturation issues over dense vegetation conditions that can be expected in the domain.}

\end{itemize} 

\subsection{Significance testing}

We sampled initiation gradients as the slope of the linear regression of surface anomalies with distance, between the CI location and a point L km downwind, where L represents the length scales of surface heterogeneity considered in this study. To compare the initiation gradients with surface conditions not linked to CI, we sampled up to eight non-initiation gradients close to the edges of the rotated surface transects extracted for each CI case. The statistical significance of the mean initiation gradient was assessed through bootstrapping \citep[refer to][]{wilks2019} against a ``superset" of all initiation and non-initiation gradients. To assess the statistical significance of the mean initiation gradient for a set of N initiation gradients, a distribution of the mean values of randomly sampled N gradients was generated from the ``superset" by repeating the sampling 1000 times with replacement. The mean value of the initiation set was deemed significant at a particular confidence level if it lay in the respective tails of the distribution of randomly sampled gradients.

\section{Results} \label{results}

\subsection{Tracked convective initiation events}

The CI tracking criteria applied to the domain during NDJFM months of 2001-2020 resulted in the identification of 77,745 CI events. Figure \ref{domain}c shows a map of CI frequency resampled to a 0.25$^{\circ}$ x 0.25$^{\circ}$ grid over the domain. Maximum CI frequency is located over the lower latitudes in northern LPB. The pattern shows a southward bulge likely associated with the northerly flow that transports moisture and momentum from lower toward higher latitudes over South America. Figure \ref{domain}d shows a diurnal peak in CI frequency in early afternoon hours between 15 UTC ($\sim$local noon) and 19 UTC ($\sim$4 pm). The normalized diurnal probability of CI occurrence in the domain reduces to less than 0.05 after 23 UTC ($\sim$8 pm local time). We consider daytime CI events as those detected between 13 UTC and 23 UTC.

\subsection{Mean surface anomalies across all satellite datasets}

\begin{figure}[!ht]
\centering
\vspace{\baselineskip}
\includegraphics[width=1\textwidth]{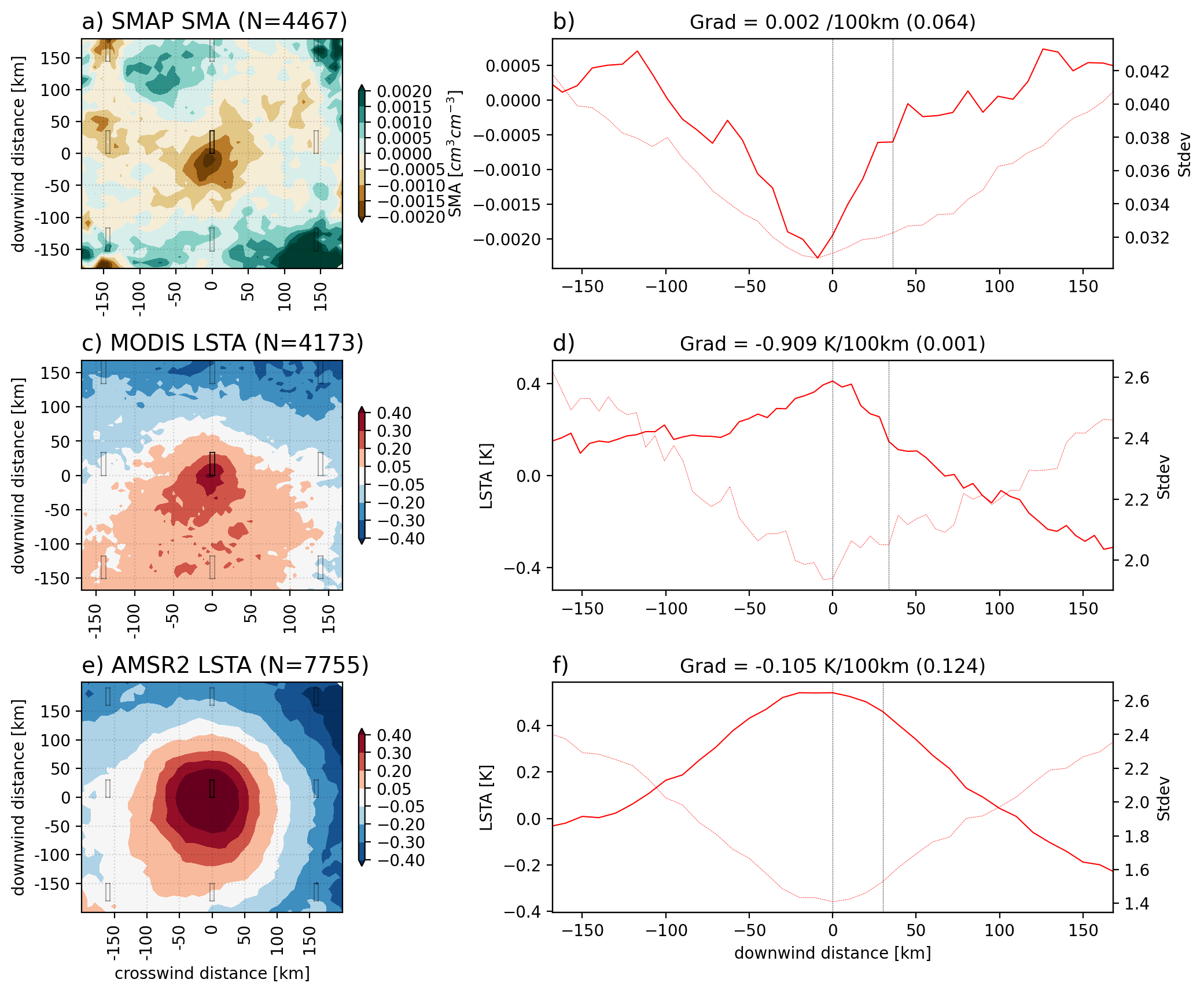}
\caption{\label{multidataset}a) Composite mean SMAP soil moisture anomaly (SMA; cm$^{3}$ cm$^{-3}$) underlying CI cases, where the SMA transects for each case were first aligned in the direction of the low-level (10 m) average wind in the two hours before the detected CI time. The center (0,0) denotes the initiation location for each case. Black rectangle (10km x 30km) at the center used to calculate the downwind SMA gradient at the initiation point. Grey rectangles were similarly used to calculate the downwind gradients over non-initiation locations for significance testing. b) Downwind cross-section of mean SMA for all CI cases (thick red curve and left ordinate) along with its standard deviation across all CI cases in the downwind direction (thin red curve and right ordinate). Dashed lines denote the extent of the box used to calculate the initiation gradient. c) Same as \textbf{a} but for MODIS land surface temperature anomaly (LSTA; K). d) Same as \textbf{b} but for MODIS LSTA. e) Same as \textbf{a} but for LSTA from AMSR2 dataset. f) Same as \textbf{b} but for AMSR2 LSTA.}
\end{figure}

Figure \ref{multidataset} presents the composite patterns of alongwind surface conditions before and directly underlying daytime CI over subtropical South America from independent sensors aboard the three different low earth orbit satellites used in our analysis. Table \ref{gradients} shows the variables used along with their sample size, the mean initiation gradient between the CI location and 30 km downwind, and the associated p-values. 

Volumetric SM anomalies from the SMAP level-3 dataset show the mean location of 4,467 CI events at the downwind end of a spatially drier surface patch (Fig. \ref{multidataset}a). The dry patch is located close to relatively moister regions contributing to enhanced heterogeneity at the surface and strong dry-wet soil boundaries at the location of CI. The one-dimensional downwind cross-section through the CI location (Fig. \ref{multidataset}b) shows that CI preferentially occurs downwind of the driest SM anomaly. A rapid increase in SM occurs on the downwind side of CI. The strength of the mesoscale gradient is quantified as the regression slope of the SMA anomalies between the ground location of CI and those approximately 30 km downwind (black rectangle in Fig. \ref{multidataset}a), similar to \citep{barton_observed_2021}. The mean downwind gradient value of +0.002 cm$^{3}$ cm$^{-3}$ (100 km)$^{-1}$ is statistically significant (p-value = 0.064) at the 90\% confidence level. The magnitude of the mean gradient is similar to the one reported by \cite{taylor_frequency_2011} over the Sahel region. This pattern suggests preferential convective initiation over dry-wet SM boundaries, at the downwind edge of the drier patch, in agreement with the hypothesis of this study. 

The analysis of MODIS LST dataset yields 4,173 CI events with valid, high quality retrievals. The mean LSTA pattern (Fig. \ref{multidataset}c) shows a spatially warmer surface anomaly upwind of the initiation location with cooler conditions downwind. Convection preferentially initiates over the downwind edge of the warm patch. The one-dimensional downwind cross-section through the CI location shows that the strongest LSTA gradient is located at and downwind of the CI location (Fig. \ref{multidataset}d). The mean gradient value of -0.909 K (100 km)$^{-1}$ over all 4,173 cases is statistically significant at the 99\% level and is stronger than similar metrics reported over the Sahel and the Tibetan Plateau.    



Analysis of the AMSR2 LST dataset serves as another validation of the surface temperature pattern. It yields valid surface transects for  7,755 daytime CI events that initiate after the GCOM-W1 overpass time, i.e. CI events between 17 UTC and 23 UTC. The general characteristics of this pattern (Fig. \ref{multidataset}e,f) match that of the MODIS LSTA analysis, although poorly resolved by comparison. The differences between the two LSTA patterns are likely due to the spatial resolution as the AMSR2 dataset is interpolated to the $\sim$10 km resolution from its native resolution of $\sim$25 km using a forward radiative transfer model. The all-weather retrieval of LST from the microwave radiometer can likely dilute the warming signal and the gradient strength due to a cool bias whenever a shallow cloud is present at the overpass time. This can be expected at the detected initiation location given that, by definition, deep convection develops at that location within minutes to a few hours.

In summary, the three satellite datasets agree on preferential CI at the downwind end of a spatially drier patch of the order of tens of kilometers, close to strong dry-wet soil boundaries. The signal from the LSTA datasets is consistent (anti-correlated) with that from the SM dataset. 


\begin{table}[!ht]
\caption{Downwind initiation gradients}
\label{gradients}
\centering
\begin{tabular}{@{} l *4c @{}}
\toprule
\multicolumn{1}{c}{\textbf{Dataset}} & \textbf{Variable} & \textbf{Initiation Gradient} & \textbf{P-value} & \textbf{Number of Cases}  \\ 
\midrule
SMAP             & SM & +0.002 (100 km)$^{-1}$       & 0.064            & 4467                    \\ 
MODIS            & LST & -0.909 K (100 km)$^{-1}$      & 0.001            & 4173                     \\ 
AMSR2            & LST & -0.105 K (100 km)$^{-1}$     & 0.124            & 7755                   \\
 \bottomrule
\end{tabular}
\end{table}

\subsection{Sensitivity to environmental conditions}

\begin{figure}[!ht]
\centering
\includegraphics[width=1.0\textwidth]{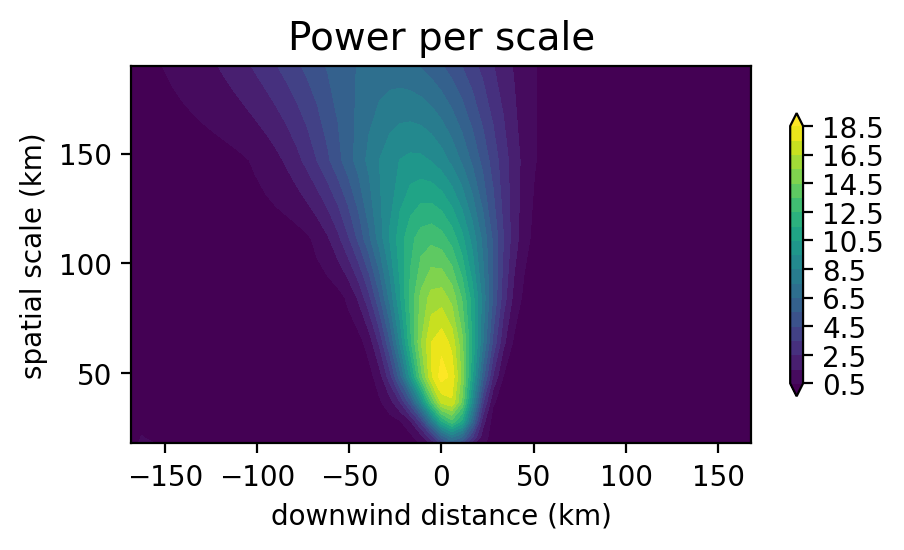}
\caption{\label{wavelets}Wavelet transform computed from the composite alongwind land surface temperature anomaly pattern on the central transect passing through the CI location in the alongwind direction, for all valid CI cases from the MODIS dataset analysis.}
\end{figure}

Convection due to mesoscale circulations induced by spatially drier soils can be modulated by environmental conditions that can be static (e.g., underlying topography) or dynamic (e.g., background wind speed) in nature. We analyze the mean spatial patterns of surface anomalies after reclassifying the CI events based on background conditions to gain insights on such interactions. Since the prestorm LSTA signal from MODIS emerges clearly above the noise and shows high statistical significance (p-value = 0.001), we employ this dataset in the subsequent analysis. The discrete spatial scales associated with the strongest surface heterogeneity were empirically determined using a wavelet analysis on the alongwind LSTA transects underlying CI. The mean wavelet transform (Fig. \ref{wavelets}) captures the strongest variability between spatial scales of 30 km and 100 km, centered close to the initiation point. The weighted amplitude of the decomposed signal is remarkably similar to the weighted wavelet transform from alongwind LSTA anomalies underlying CI over the Sahel region \citep[Fig. 2b of][]{taylor_frequency_2011}. Here, we present the downwind initiation gradients for the lower (30 km) and upper (100 km) ends of the dominant wavelength spectrum.


Table \ref{gradients_reclassified} shows the mean initiation gradients from the MODIS dataset at 30 km and 100 km downwind of the initiation point respectively, after resampling the CI events based on various thresholds of a) seasonality, represented through bimonthly periods, b) topographic complexity, i.e. the local standard deviation of elevation in the 40 km-by-40 km region centered at the initiation point, c) background low-level (10 m) wind, d) background CAPE and e) background CIN in the two hours before the detected initiation time averaged over the 1$^{\circ}$-by-1$^{\circ}$ box centered at the initiation location; and f) background vegetation, represented through monthly EVI at the CI location. The statistical significance of the mean initiation gradient at a particular length scale for each of the sub-classes was assessed against the ``superset" of initiation and non-initiation gradients as done for Figure \ref{multidataset}c but for the respective length scale.


\begin{table}[]
\caption{Downwind initiation gradients from MODIS LSTA dataset at 30 km and 100 km length scales along with the p-values from permutation tests shown in parenthesis. The criteria or thresholds for the different sub-classes are shown in italics. Bold quantities denote statistical significance at the 95\% confidence level.}
\label{gradients_reclassified}
\resizebox{\textwidth}{!}{
\begin{tabular}{@{}c|cccc|cccc@{}}
\toprule
\textbf{Parameter}                                                              & \multicolumn{4}{c|}{\textbf{\begin{tabular}[c]{@{}c@{}}30 km gradient \\ {[}K (100 km)$^{-1}${]}\end{tabular}}}                                                                                                                                                               & \multicolumn{4}{c}{\textbf{\begin{tabular}[c]{@{}c@{}}100 km gradient \\ {[}K (100 km)$^{-1}${]}\end{tabular}}}                                                                                                                                                               \\ \midrule
\multirow{2}{*}{Months}                                                         & \textit{ND}                                                       & \textit{DJ}                                                       & \textit{JF}                                                       & \textit{FM}                                                       & \textit{ND}                                                       & \textit{DJ}                                                       & \textit{JF}                                                       & \textit{FM}                                                       \\ \cmidrule(l){2-9} 
                                                                                & \textbf{\begin{tabular}[c]{@{}c@{}}-0.934\\ (0.007)\end{tabular}} & \textbf{\begin{tabular}[c]{@{}c@{}}-0.756\\ (0.015)\end{tabular}} & \textbf{\begin{tabular}[c]{@{}c@{}}-0.892\\ (0.003)\end{tabular}} & \textbf{\begin{tabular}[c]{@{}c@{}}-0.995\\ (0.002)\end{tabular}} & \textbf{\begin{tabular}[c]{@{}c@{}}-0.543\\ (0.001)\end{tabular}} & \textbf{\begin{tabular}[c]{@{}c@{}}-0.479\\ (0.001)\end{tabular}} & \textbf{\begin{tabular}[c]{@{}c@{}}-0.451\\ (0.001)\end{tabular}} & \textbf{\begin{tabular}[c]{@{}c@{}}-0.485\\ (0.001)\end{tabular}} \\ \midrule
\multirow{2}{*}{\begin{tabular}[c]{@{}c@{}}TC\\ {[}m{]}\end{tabular}}           & \textit{2-36}                                                     & \multicolumn{2}{c}{\textit{36-89}}                                                                                                    & \textit{89-1074}                                                  & \textit{2-36}                                                     & \multicolumn{2}{c}{\textit{36-89}}                                                                                                    & \textit{89-1074}                                                           \\ \cmidrule(l){2-9} 
                                                                                & \textbf{\begin{tabular}[c]{@{}c@{}}-0.867\\ (0.008)\end{tabular}} & \multicolumn{2}{c}{\textbf{\textbf{\begin{tabular}[c]{@{}c@{}}-0.985\\ (0.001)\end{tabular}}}}                                                 & \textbf{\begin{tabular}[c]{@{}c@{}}-0.867\\ (0.01)\end{tabular}}          & \textbf{\begin{tabular}[c]{@{}c@{}}-0.471\\ (0.001)\end{tabular}} & \multicolumn{2}{c}{\textbf{\begin{tabular}[c]{@{}c@{}}-0.337\\ (0.025)\end{tabular}}}                                                          & \textbf{\begin{tabular}[c]{@{}c@{}}-0.602\\ (0.001)\end{tabular}}          \\ \midrule
\multirow{2}{*}{\begin{tabular}[c]{@{}c@{}}EVI\\ {[}ratio{]}\end{tabular}}      & \textit{0.1-0.4}                                                  & \multicolumn{2}{c}{\textit{0.4-0.5}}                                                                                                  & \textit{0.5-0.9}                                                  & \textit{0.1-0.4}                                                  & \multicolumn{2}{c}{\textit{0.4-0.5}}                                                                                                  & \textit{0.5-0.9}                                                          \\ \cmidrule(l){2-9} 
                                                                                & \textbf{\begin{tabular}[c]{@{}c@{}}-1.367\\ (0.001)\end{tabular}} & \multicolumn{2}{c}{\begin{tabular}[c]{@{}c@{}}-0.622\\ (0.069)\end{tabular}}                                                          & \begin{tabular}[c]{@{}c@{}}-0.676\\ (0.055)\end{tabular}          & \textbf{\begin{tabular}[c]{@{}c@{}}-0.766\\ (0.001)\end{tabular}} & \multicolumn{2}{c}{\begin{tabular}[c]{@{}c@{}}-0.248\\ (0.168)\end{tabular}}                                                          & \textbf{\begin{tabular}[c]{@{}c@{}}-0.365\\ (0.029)\end{tabular}}          \\ \midrule
\multirow{2}{*}{\begin{tabular}[c]{@{}c@{}}CAPE\\ {[}J/kg{]}\end{tabular}}      & \textit{0-600}                                                    & \multicolumn{2}{c}{\textit{600-1500}}                                                                                                 & \textit{1500-4000}                                                & \textit{0-600}                                                    & \multicolumn{2}{c}{\textit{600-1500}}                                                                                                 & \textit{1500-4000}                                                         \\ \cmidrule(l){2-9} 
                                                                                & \begin{tabular}[c]{@{}c@{}}-0.548\\ (0.147)\end{tabular}          & \multicolumn{2}{c}{\textbf{\begin{tabular}[c]{@{}c@{}}-1.014\\ (0.001)\end{tabular}}}                                                 & \textbf{\begin{tabular}[c]{@{}c@{}}-1.316\\ (0.001)\end{tabular}}          & \textbf{\begin{tabular}[c]{@{}c@{}}-0.324\\ (0.027)\end{tabular}} & \multicolumn{2}{c}{\textbf{\begin{tabular}[c]{@{}c@{}}-0.578\\ (0.001)\end{tabular}}}                                                 & \textbf{\begin{tabular}[c]{@{}c@{}}-0.566\\ (0.001)\end{tabular}}          \\ \midrule
\multirow{2}{*}{\begin{tabular}[c]{@{}c@{}}CIN\\ {[}J/kg{]}\end{tabular}}       & \textit{0-100}                                                    & \multicolumn{2}{c}{\textit{100-250}}                                                                                                  & \textit{250-1000}                                                 & \textit{0-100}                                                    & \multicolumn{2}{c}{\textit{100-250}}                                                                                                  & \textit{250-1000}                                                          \\ \cmidrule(l){2-9} 
                                                                                & \textbf{\begin{tabular}[c]{@{}c@{}}-1.052\\ (0.001)\end{tabular}} & \multicolumn{2}{c}{\textbf{\begin{tabular}[c]{@{}c@{}}-0.934\\ (0.001)\end{tabular}}}                                                 & \textbf{\begin{tabular}[c]{@{}c@{}}-0.885\\ (0.016)\end{tabular}}          & \begin{tabular}[c]{@{}c@{}}-0.313\\ (0.096)\end{tabular}          & \multicolumn{2}{c}{\textbf{\begin{tabular}[c]{@{}c@{}}-0.51\\ (0.001)\end{tabular}}}                                                  & \textbf{\begin{tabular}[c]{@{}c@{}}-0.644\\ (0.001)\end{tabular}} \\ \midrule
\multirow{2}{*}{\begin{tabular}[c]{@{}c@{}}Wind speed\\ {[}m/s{]}\end{tabular}} & \textit{0-1.5}                                                    & \multicolumn{2}{c}{\textit{1.5-2.5}}                                                                                                  & \textit{2.5-10.0}                                                 & \textit{0-1.5}                                                    & \multicolumn{2}{c}{\textit{1.5-2.5}}                                                                                                  & \textit{2.5-10.0}                                                          \\ \cmidrule(l){2-9} 
                                                                                & \textbf{\begin{tabular}[c]{@{}c@{}}-1.032\\ (0.001)\end{tabular}} & \multicolumn{2}{c}{\textbf{\begin{tabular}[c]{@{}c@{}}-1.012\\ (0.001)\end{tabular}}}                                                 & \begin{tabular}[c]{@{}c@{}}-0.613\\ (0.104)\end{tabular}            & \textbf{\begin{tabular}[c]{@{}c@{}}-0.371\\ (0.012)\end{tabular}} & \multicolumn{2}{c}{\textbf{\begin{tabular}[c]{@{}c@{}}-0.669\\ (0.001)\end{tabular}}}                                                 & \textbf{\begin{tabular}[c]{@{}c@{}}-0.347\\ (0.043)\end{tabular}} \\ \bottomrule
\end{tabular}}
\end{table}

\subsubsection{Seasonality}

\begin{figure}[!h]
\centering
\vspace{\baselineskip}
\includegraphics[width=1.0\textwidth]{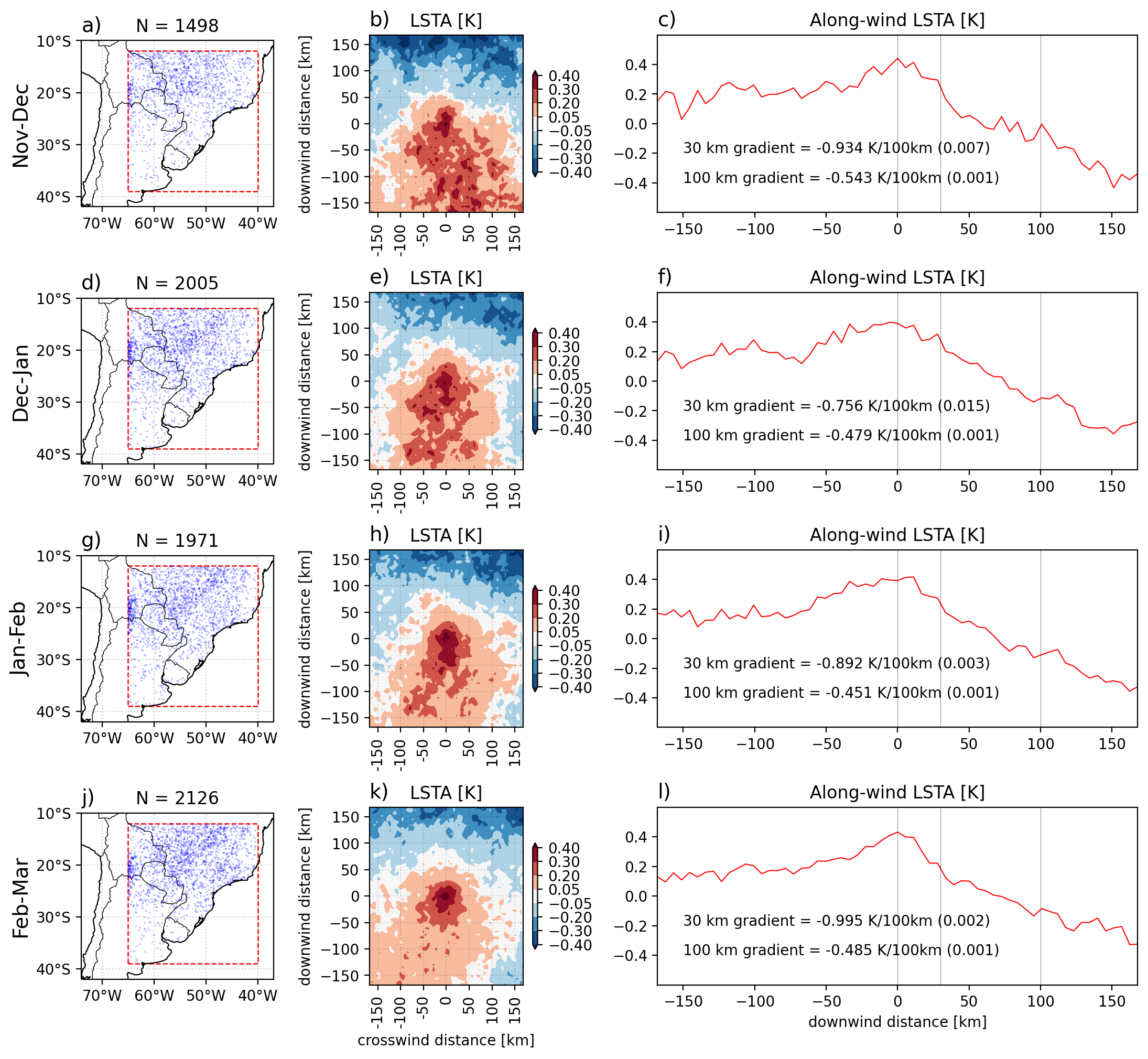}
\caption{\label{seasonality_main}Location of detected CI events (first column), composite mean MODIS LSTA (K; shading; middle column) and downwind cross-section of mean LSTA passing through the CI location (last column) for CI cases classified based on the bimonthly period in which they occur. a-c) November-December, d-f) December-January, g-i) January-February, and j-l) February-March. The LSTA data for each case is first aligned in the direction of the low-level (10 m) average wind in the two hours before the detected initiation time. For the middle column, the center (0,0) denotes the initiation location for each case. For the last column, dashed lines denote the extent of the boxes used to calculate the initiation gradients at the 30 km and the 100 km length scales.}
\end{figure}



The location of CI events is similarly distributed over the domain during all bimonthly periods (left column of Fig. \ref{seasonality_main}). These results show a strong negative LSTA gradient underlying CI events throughout the season over both length scales (Fig. \ref{seasonality_main} and Table \ref{gradients_reclassified}). The spatial LSTA pattern shows slightly stronger negative anomalies over a larger region earlier in the season (November through January) compared to later in the season (January through March) when the negative anomalies are confined to a smaller patch. The initiation gradient at 100 km length scale is also strongest earlier in the season (Nov-Dec = -0.543 K/100 km) and weakens as the season progresses (Table \ref{gradients_reclassified}). Contrary to expectations, we do not find strong evidence for intra-seasonal variability of the hypothesized negative feedback.



\subsubsection{Topographic complexity}

\begin{figure}[!h]
\centering
\vspace{\baselineskip}
\includegraphics[width=1.0\textwidth]{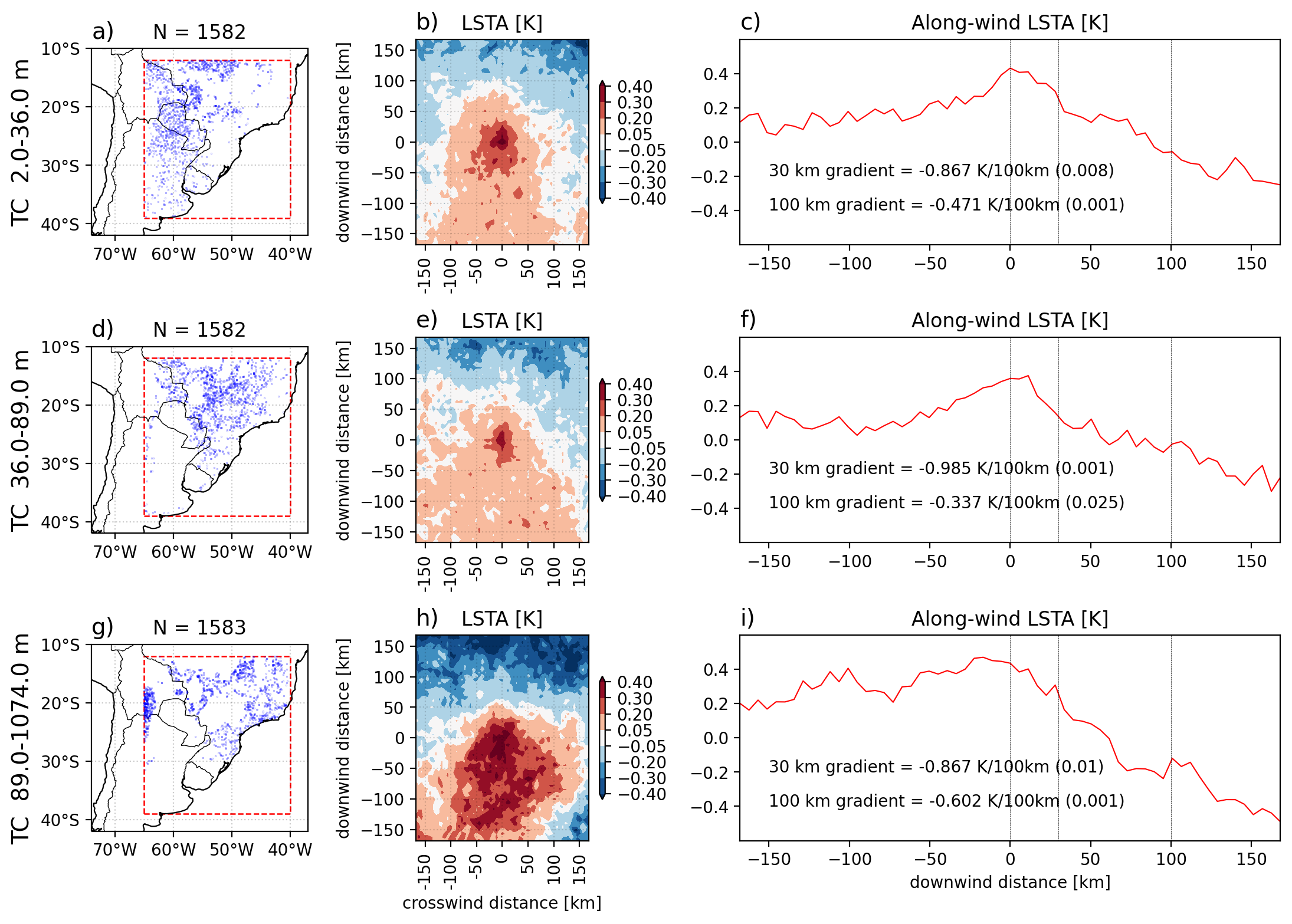}
\caption{\label{topo_main} Same as Fig. \ref{seasonality_main} but for CI cases classified based on terciles of topographic complexity over which they occur. a-c) 2 to 36 m , d-f) 36 to 89 m, and g-i) greater than 89 m.}
\end{figure}

Topography can induce upslope circulations due to efficient heating on elevated surfaces as well as through mechanical lifting of unstable near-surface air, leading to convective onset.  We explore the differences in the downwind initiation gradients for CI cases re-classified into terciles of topographic complexity, i.e. the local standard deviation of elevation in a 40 km-by-40 km region centered at the ground location of CI (Fig. \ref{topo_main}). The size of the region to define TC reflects a $\beta$-mesoscale region associated with convective onset where the mechanical and buoyancy effects of topography in forcing upslope flow are effective. 


Figure \ref{topo_main} shows the CI locations, spatial patterns of surface anomalies and the associated central transects for three non-overlapping TC classes. The thresholds for the TC bins are determined to approximately divide the CI events into terciles. These results show preferential CI over strong negative LSTA gradients across all terciles of TC at the CI location (Table \ref{gradients_reclassified}). 

\begin{figure}[!h]
\centering
\vspace{\baselineskip}
\includegraphics[width=1.0\textwidth]{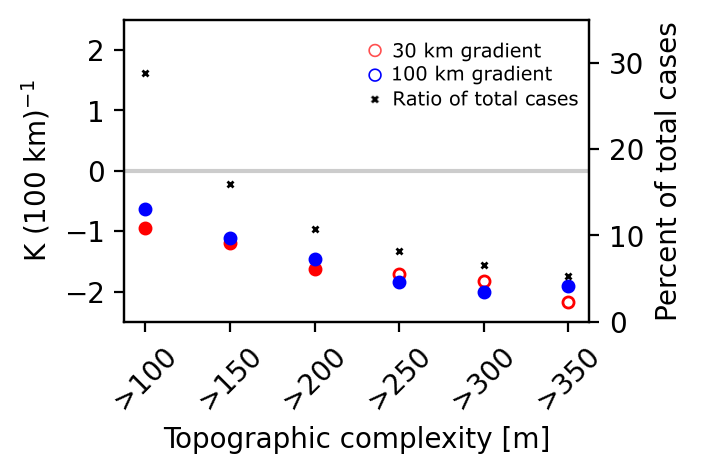}
\caption{\label{topo_highclasses}Downwind LSTA gradients at the initiation point from the MODIS dataset that occur over a minimum topographic complexity indicated by the x-axis labels. Filled circles denote statistically significant gradient at the 99\% confidence level. Black crosses indicate the ratio of total CI events associated with each sub-class.}
\end{figure}

The spatial pattern for the highest TC tercile shows negative LSTAs over a larger region. At the 100 km scale, the gradient strength for the highest TC tercile (-0.602 K/100 km) is approximately twice that of the lower tercile (-0.337 K/100 km). This suggests that larger scale LSTA gradients could be important for convection over higher terrain. We further investigate the highest TC tercile (TC $>$ 89 m) by incrementally increasing the minimum TC value, e.g. TC $>$100 m, 150 m, and so on. Figure \ref{topo_highclasses} shows the downwind initiation gradients, their statistical significance at the 99\% level and the ratio of total CI cases (out of 4173) associated with these new sub-classes within the highest tercile. The 30 km gradient is significant for TC sub-classes up to 250 m \citep[similar to][]{barton_observed_2021}. We find increasing strength of the gradient with increasing TC. 

\begin{figure}[!h]
\centering
\includegraphics[width=1.0\textwidth]{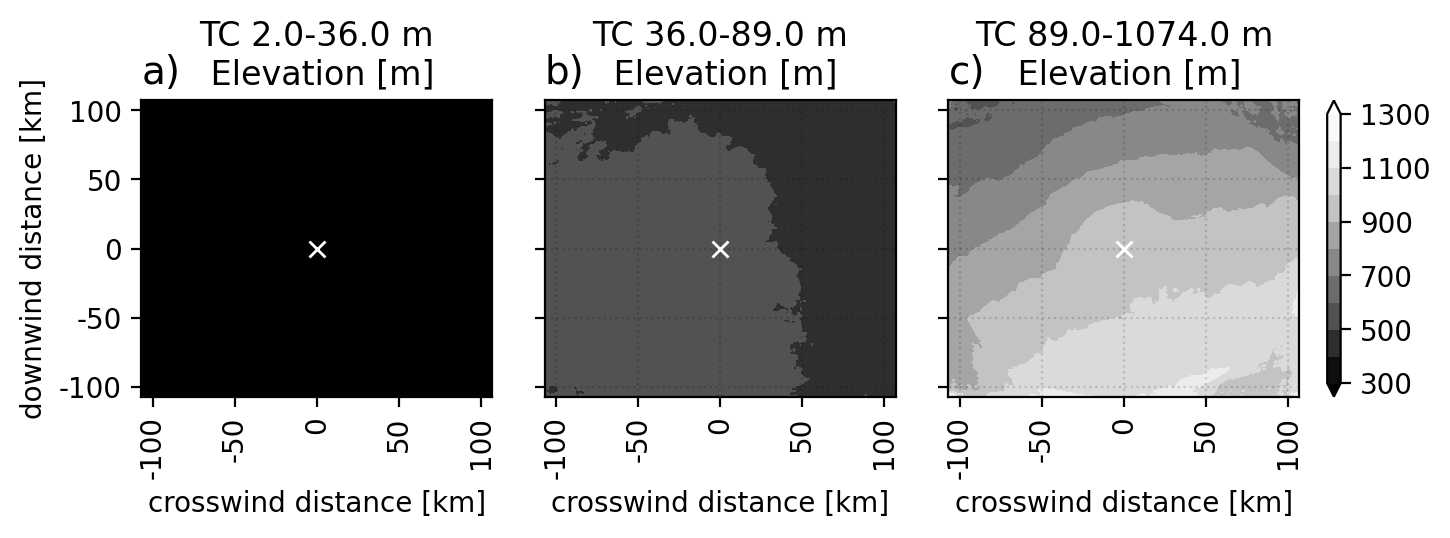}
\caption{\label{topo_elev} Composite mean elevation (m; shading) underlying CI cases, where the elevation data for each case is first aligned in the direction of the low-level (10 m) in a similar way as Fig. \ref{topo_main}, for CI cases classified based on terciles of topographic complexity over which they occur. a) 2 to 36 m , b) 36 to 89 m, and c) greater than 89 m. Note the alongwind decline in elevation for the highest tercile.}
\end{figure}

Figure \ref{topo_elev} shows the composite pattern of the alongwind elevation transects centered at the CI locations. While the pattern for the lowest TC tercile shows a relatively flat region, the pattern for the highest tercile shows steep alongwind decline in elevation. The CI cases in the highest tercile occur when the induced upslope circulation due to elevated warming over a spatially drier patch opposes the downslope background flow. These observed patterns are consistent with the modeling study of \cite{Imamovic_orography_2017} who found that the superposition of mountain-valley circulation and the SM heterogeneity-induced flow due to an idealized small mountain can lead to increased rainfall accumulation because mountains that are drier than the nearby low-lying regions can induce stronger circulations.



\subsubsection{Background vegetation index}

\begin{figure}[!h]
\centering
\includegraphics[width=1.0\textwidth]{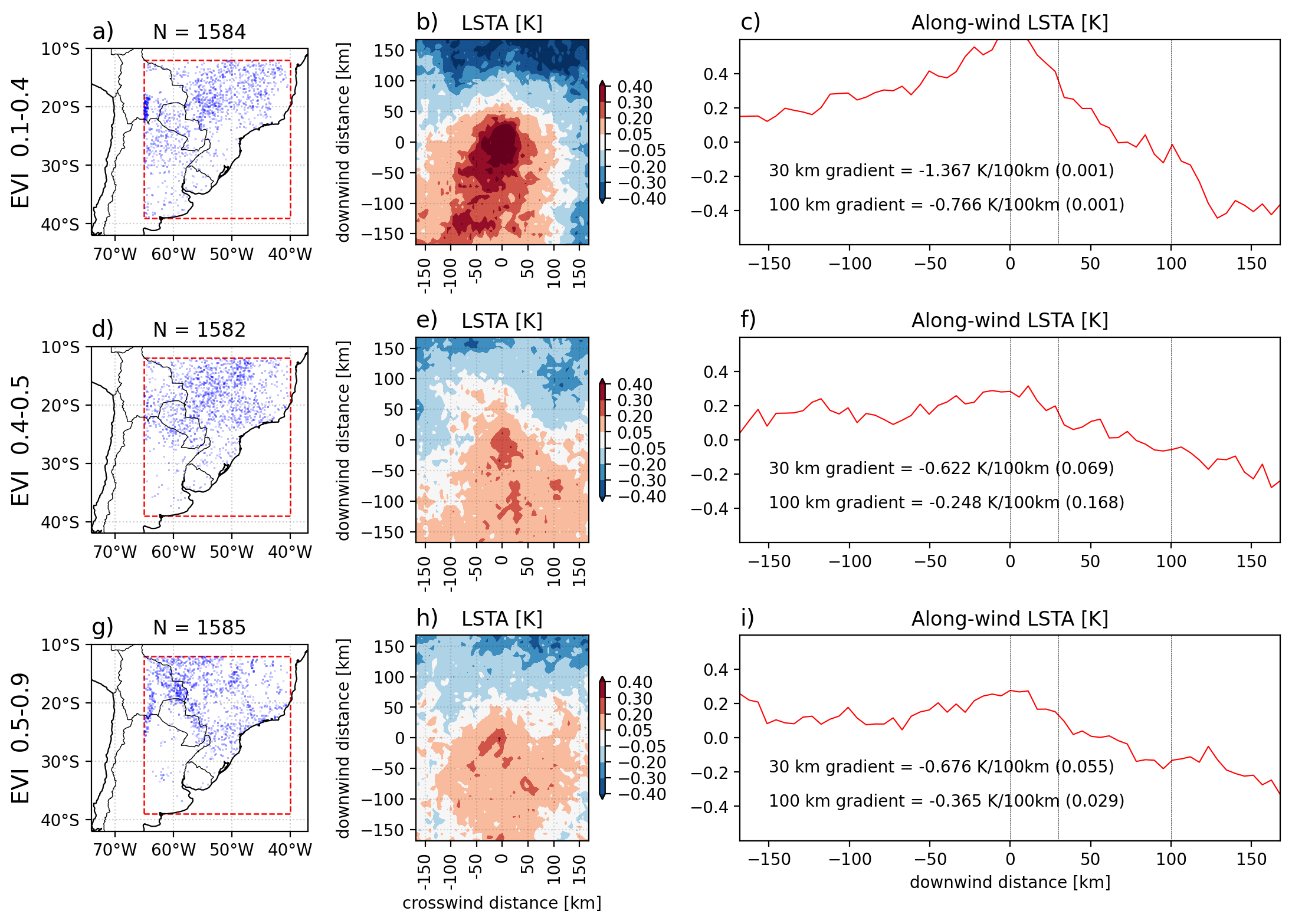}
\caption{\label{veg_main}Same as Fig. \ref{seasonality_main} but for CI cases classified based on terciles of background EVI. a-c) Less than 0.4, d-f) 0.4 to 0.5, and g-i) greater than 0.5.}
\end{figure}


The idealized mechanism for CI over spatially drier bare-soils can be modulated by the intermediary role of vegetation that can affect the exchange of moisture, energy and momentum with the above-canopy boundary layer. Subtropical South America exhibits a variety of land cover types - evergreen and deciduous broadleaf forests, savannas and woody savannas, grasslands and croplands. To test the sensitivity of convective onset to vegetation density, we employ the MODIS EVI to re-classify the CI cases approximately into terciles of low ($<$0.4), moderate (0.4-0.5) and high EVI ($>$0.5). Figure \ref{veg_main} shows a clear signal of statistically significant (99\% level), negative LSTA gradient underlying CI cases in the lowest EVI tercile (EVI $<$ 0.4). These regions correspond to savanna, grassland and cropland land cover classes mainly over northern Argentina, Paraguay and southern Brazil. This sensitivity is clear and consistent across both length scales (Table \ref{gradients_reclassified}). Results for higher EVI classes show negative LSTA gradients that are relatively weaker compared to sparse vegetation cases.

\subsubsection{Convective potential and inhibition}

\begin{figure}[!h]
\centering
\includegraphics[width=1.0\textwidth]{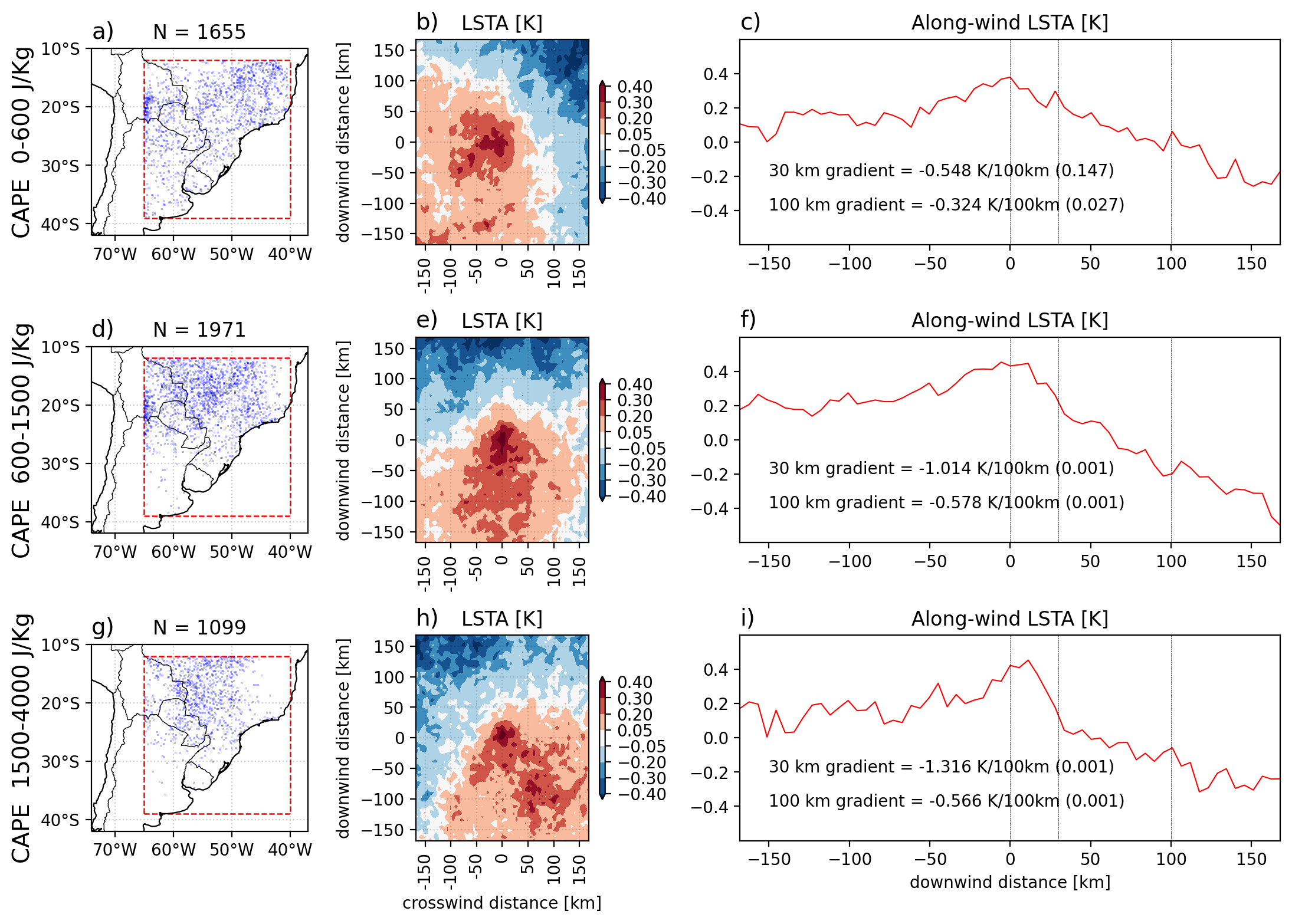}
\caption{\label{cape_main}Same as Fig. \ref{seasonality_main} but for CI cases classified based on terciles of background CAPE in the two hours before initiation. a-c) Less than 600 J/kg , d-f) 600 to 1500 J/kg, and g-i) greater than 1500 J/kg.}
\end{figure}

CAPE provides a useful measure of the maximum possible intensity of surface-based convection. At convective-scales, the structure and length scale of the surface heterogeneity can strongly influence the ability of mesoscale flows to concentrate CAPE within regions where the likelihood of stronger thunderstorm is locally maximized \citep{Avissar_shallowclouds_1996, Pielke_vegetation_2001}. Notably, the estimates of CAPE from ERA5 dataset at 0.25$^{\circ}$ resolution used in this study are - a) strongly influenced by model physics rather than observed profiles, and b) likely represent the domain mean conditions, i.e. they do not adequately capture the effects of surface heterogeneity at convective scales. 

While CAPE provides a good measure of the potential for strong convection, breaking the energy barrier below the level of free convection is critical to initiate convection. CIN is a strong determinant of the location of CI when expansive regions may have similarly positive CAPE. Unlike CAPE, changes in surface temperature have a greater effect on the CIN than corresponding changes in low-level humidity that lead to the same change in moist static energy within a well-mixed PBL \citep{Trier_CI_2003}. Spatially drier/warmer soils are ideally situated to help reduce the CIN locally before the development of convection.

Figure \ref{cape_main} shows that the highest CAPE cases are limited to the low-lying regions of southern Amazon and the Gran Chaco in northern LPB. For both length scales, the strength of the negative downwind gradient increases with increasing CAPE (Table \ref{gradients_reclassified}). The strongest 30 km gradients are observed for cases with high background CAPE ($>$ 1500 J/kg). This suggests that convective forcing from small scale gradients is effective when the lower troposphere outside the boundary layer is more conducive to convection. 


\begin{figure}[!h]
\centering
\includegraphics[width=1.0\textwidth]{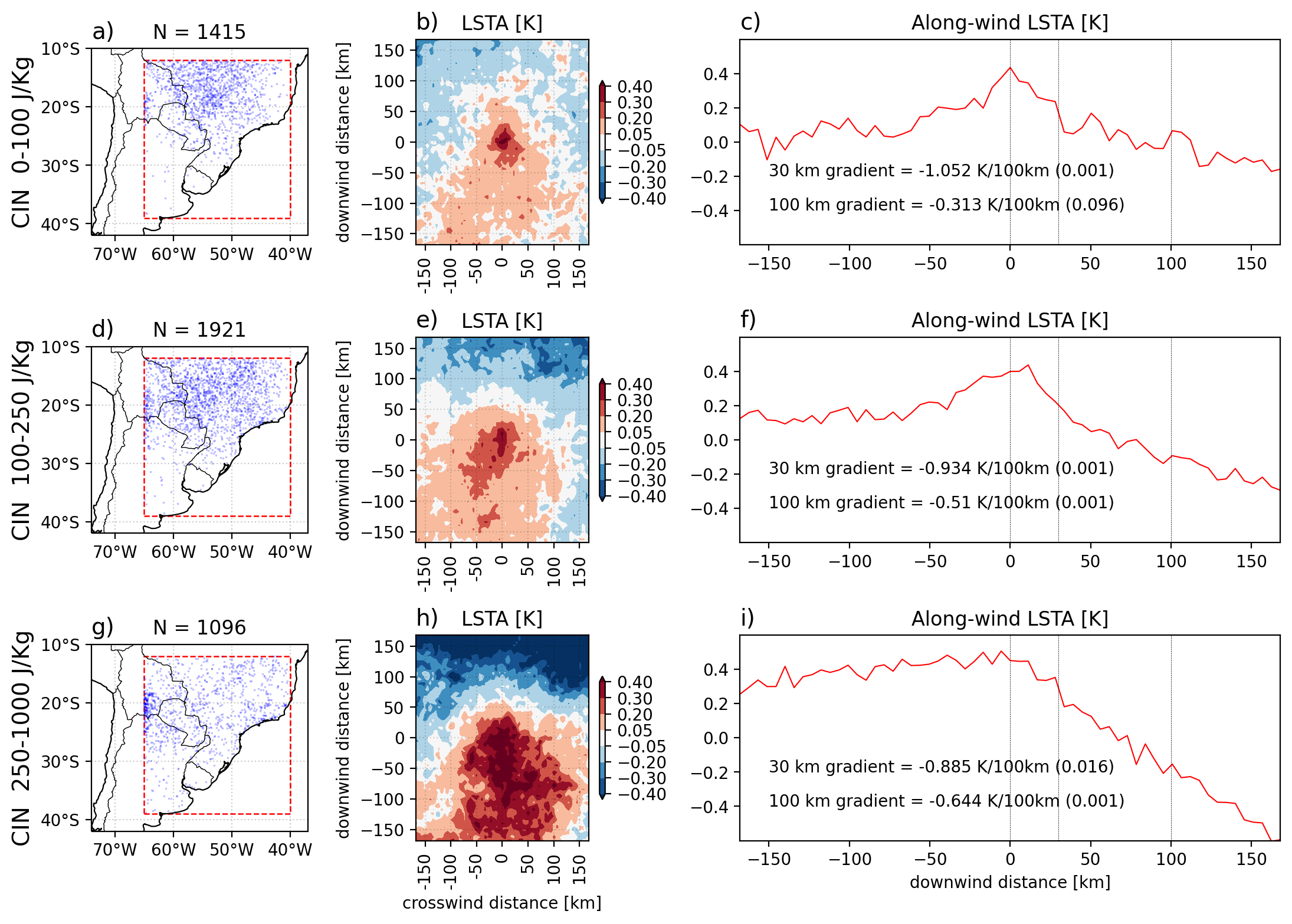}
\caption{\label{cin_main}Same as Fig. \ref{seasonality_main} but for CI cases classified based on terciles of background CIN in the two hours before initiation. a-c) Less than 100 J/kg , d-f) 100 to 250 J/kg, and g-i) greater than 250 J/kg.}
\end{figure}

The statistically significant gradient at the 100 km scale in the lowest CAPE tercile suggests that the surface heterogeneity of a larger length scale can play an important role in initiating convection during less favorable synoptic conditions. This is evident in the contrasting relationship of the downwind initiation gradient with increasing CIN at the CI location (Table \ref{gradients_reclassified}). While strong LSTA gradients at the 30 km length scale are effective in initiating convection in convectively favorable environments up to moderate CIN conditions ($<$ 250 J/kg), the strength and significance of the 100 km gradients consistently increases with increasing CIN. Figure \ref{cin_main} shows a stronger and more expansive pattern of warmer anomalies with increasing thresholds of background CIN. Convection in high CIN environments is insensitive to smaller scale SMA/LSTA gradients. However, when present over a larger scale, these gradients can generate persistent mesoscale circulations that a) increase the CAPE locally through convergence, and b) reduce the CIN locally through surface warming to overcome the larger inhibition.



\subsubsection{Low-level wind} \label{wind_section}

\begin{figure}[!h]
\centering
\includegraphics[width=1.0\textwidth]{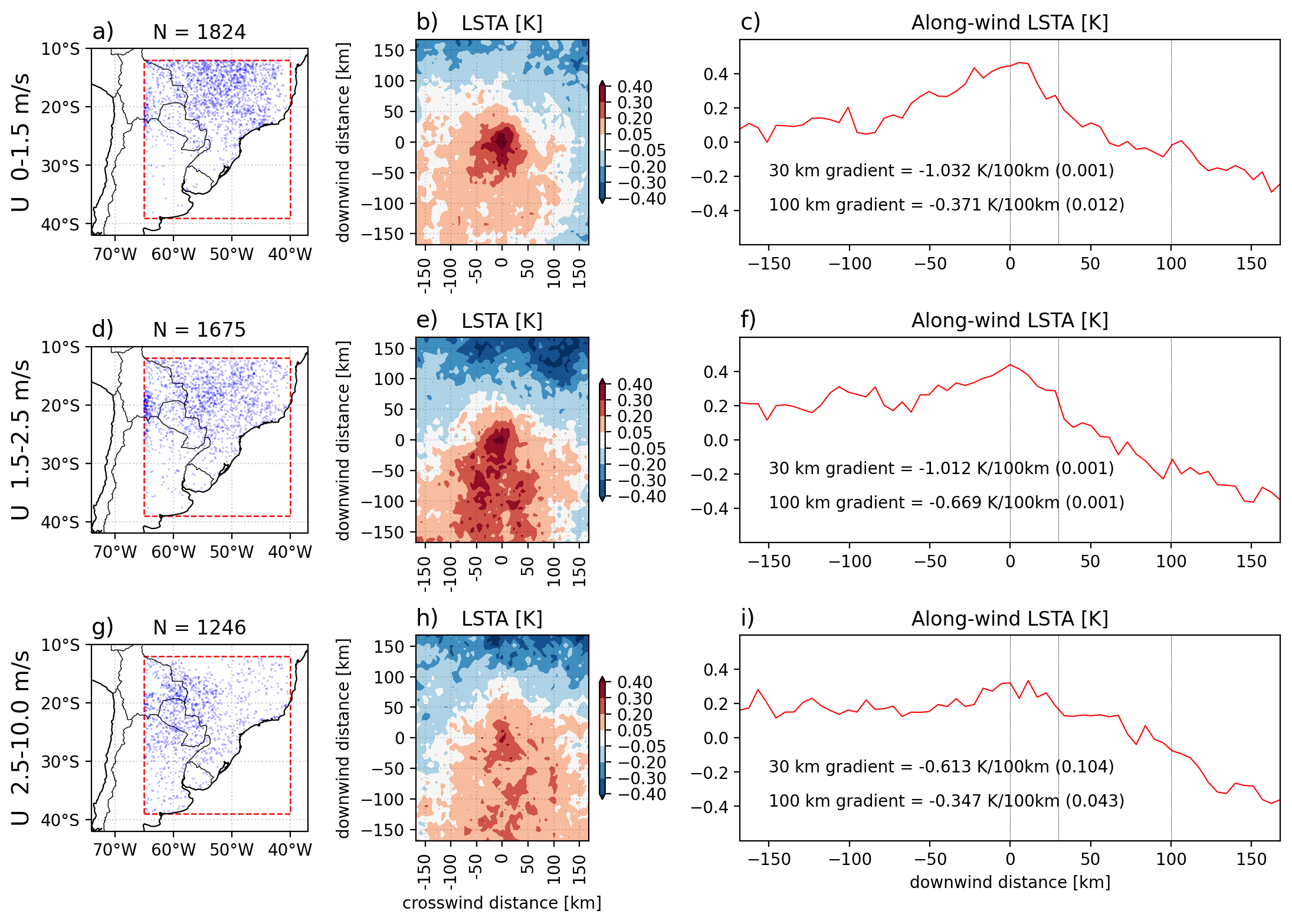}
\caption{\label{wind_main}Same as Fig. \ref{seasonality_main} but for CI cases classified based on terciles of background 10 m wind speed in the two hours before initiation. a-c) Less than 1.5 m/s , d-f) 1.5 to 2.5 m/s, and g-i) greater than 2.5 m/s.}
\end{figure}

Low-level wind speed (10 m) is used as a proxy for background flow that can lead to enhanced horizontal mixing and affect the persistence of surface-induced variability propagating into the daytime boundary layer. The distribution of meridional 10 m wind speeds before the detected CI events at the initiation location (Supplementary Fig. S1) shows higher probability for convective onset during northerly flow.

We find significantly stronger negative LSTA gradients at 30 km scale underlying CI for background wind speeds less than 2.5 m/s (Table \ref{gradients_reclassified}). This is in agreement with results from large-eddy simulations that show a suppression of the effect of surface heterogeneity on CI when wind speeds are greater than 2.5 m/s on length scales up to 40 km \citep[and reference therein]{lee_effect_2019}. Similar sensitivity to background wind speed was found in \cite{barton_observed_2021} over the Tibetan Plateau for CI cases with underlying TC less 300 m. 

The location of initiations (left column of Fig. \ref{wind_main}) shows a southwestward migration of CI locations with increasing background wind speeds. Stronger low-level flow in this region is associated with the southward penetration of the SALLJ. 

The spatial patterns of LSTA anomalies underlying convection show a distinct, spatial gradient on the order of tens of kilometers for wind speeds less than 2.5 m/s, however, the length scale of the upwind warmer area seems to increase with increasing wind speed (middle column in Fig. \ref{wind_main}). This is reflected in the statistically significant downwind gradients at 100 km scale for all three terciles of wind speed (Table \ref{gradients_reclassified}). Under conditions of strong background wind, the atmosphere is insensitive to spatial variability of smaller length scale. However, convection is still favored over larger scale negative LSTA gradients, consistent with \cite{barton_observed_2021}.

\subsection{What determines the sign of SM-PPT feedbacks?}

\begin{figure}[!h]
\centering
\vspace{\baselineskip}
\includegraphics[width=1.0\textwidth]{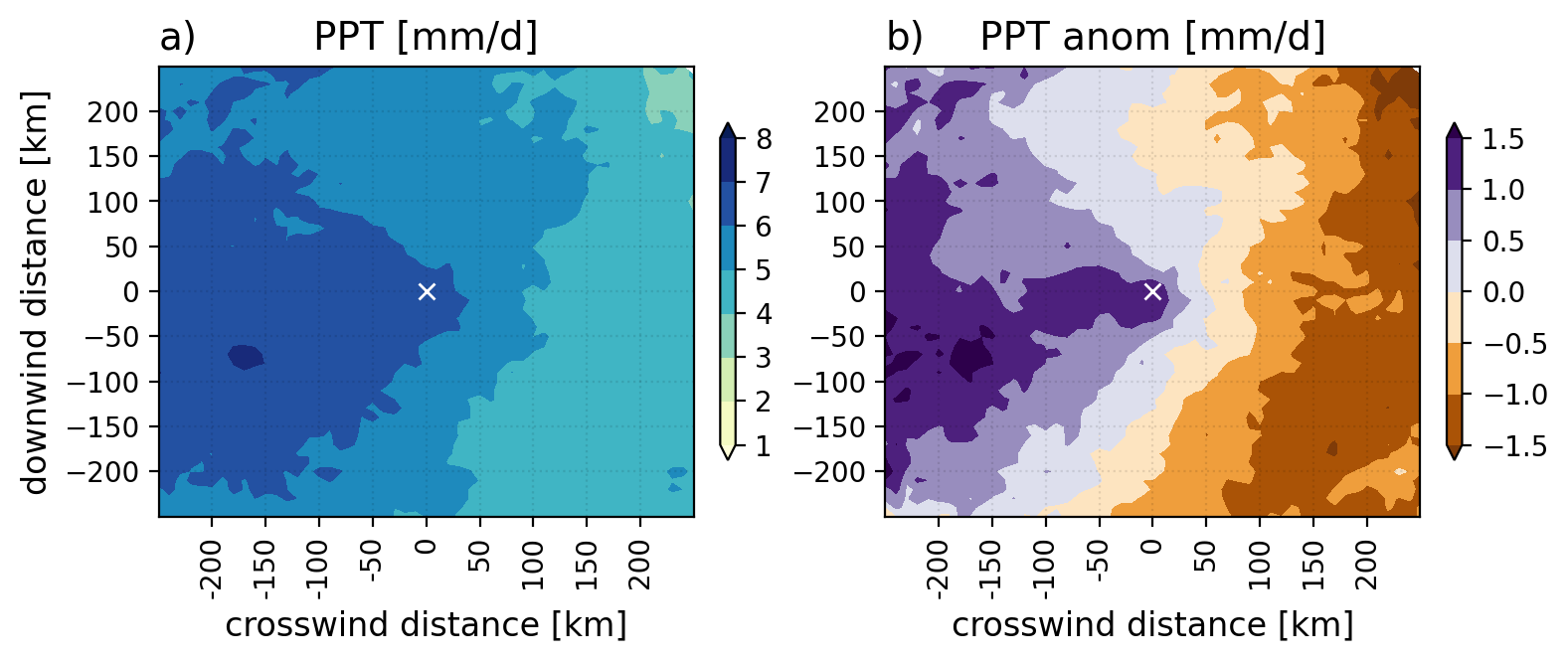}
\caption{\label{ppt}a) Composite mean precipitation (PPT) rate and b) PPT anomaly averaged over 24 hours following the initiation of convection for all valid cases in the MODIS analysis set (N = 4173). The PPT and PPT anomaly data for each case is first aligned in the direction of the low-level (10 m) average wind. The white cross at (0,0) denotes the initiation location for each case.}
\end{figure}

We analyzed the composite PPT and PPT anomaly patterns over 24 hours following the initiation time of each CI event using the IMERG half-hourly high quality microwave-only dataset (Fig. \ref{ppt}). We rotated each PPT transect using the same low level wind direction used in the respective MODIS transects in Fig. \ref{multidataset}c. Figure \ref{ppt} shows that the rainfall-producing storms preferentially translate leftward of the downwind direction in the rotated composite pattern. Since the majority of storms initiate under the climatological northerly low-level flow, the composite pattern implies eastward propagation of rain-producing storms in a geographic coordinate system (Supplementary Fig. S2). Similar eastward progression of extreme rain features in subtropical South America has previously been shown by \cite{janowiak_cmorph_2005} and \cite{rasmussen_contribution_2016} who analyzed mesoscale convective systems over their lifecycle of several hours. The composite pattern of PPT and PPT anomaly in Fig. \ref{ppt} along with Fig. \ref{multidataset} show that - a) afternoon storm initiation is favored through the convergence produced by the induced mesoscale circulation against the background flow over strong dry-wet SM boundaries, and b) the location of maximum PPT can vary between the initiation point and hundreds of kilometers eastward of the CI location. The translation of maximum PPT confounds the sign of the convective-scale SM-PPT feedback if collocated values of SM and PPT are considered. This demonstrates a fundamental issue in defining convective scale SM-PPT feedbacks similar to ``local" SM-PPT feedbacks that have previously been derived from coarse-scale climate model outputs.  

\begin{figure}[!h]
\centering
\vspace{\baselineskip}
\includegraphics[width=1.0\textwidth]{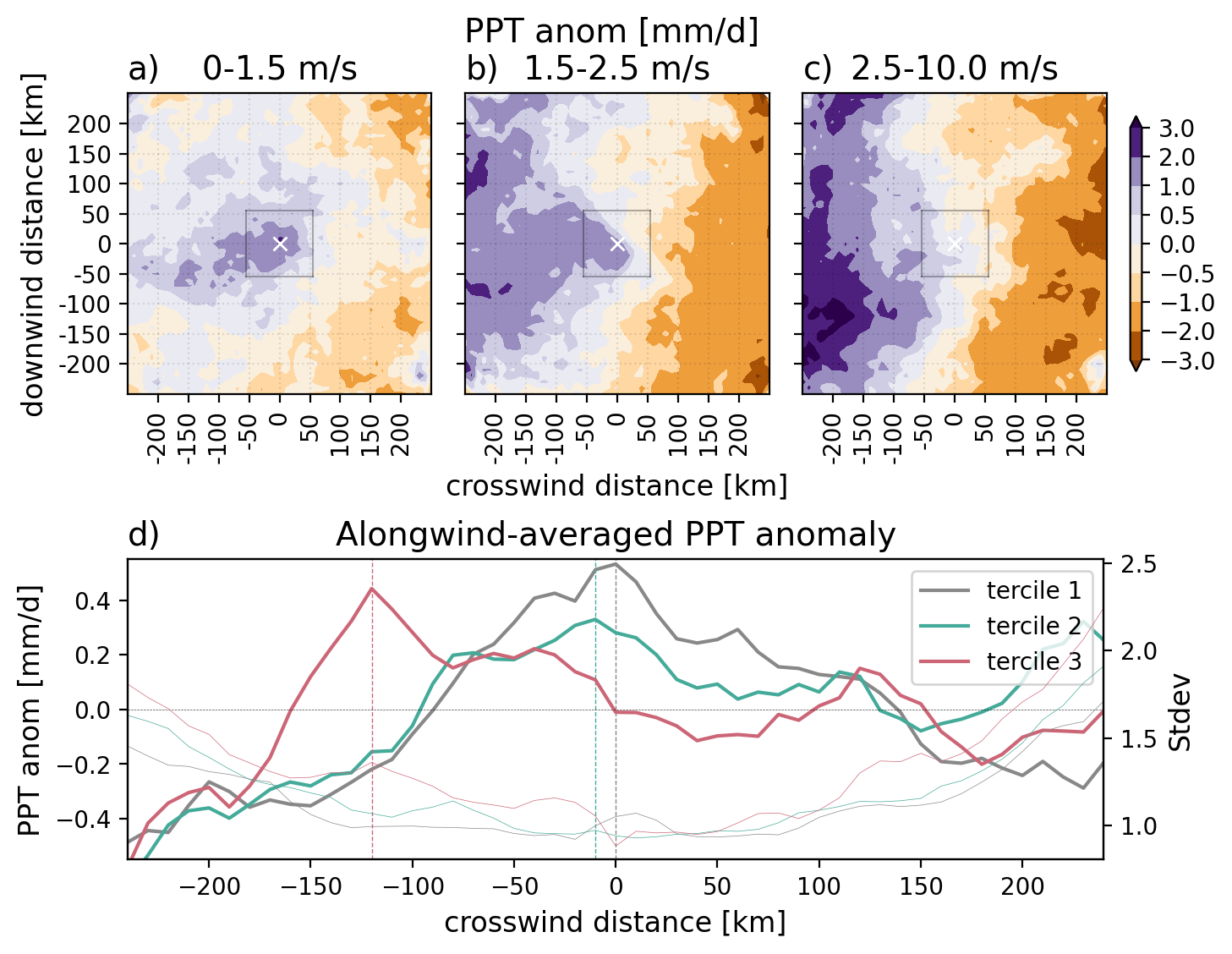}
\caption{\label{ppt_wind}Composite alongwind PPT anomaly (mm/d) averaged over 24 hours following the initiation of convection for all valid cases from the MODIS dataset classified approximately based on terciles of background wind (similar to Fig. \ref{wind_main}) shown as a) Less than 1.5 m/s , b) 1.5 to 2.5 m/s, and c) greater than 2.5 m/s. Grey box indicates the area used for computing spatial correlation between LSTA and PPT anomalies. d) Mean PPT anomaly (thick curves) and standard deviation (thin curves) for each tercile of CI events after averaging in the alongwind direction over the domain shown in (a-c). Dashed vertical lines show the peak alongwind-averaged PPT in the crosswind direction for each tercile.}
\end{figure}

We postulate that the sign of the SM-PPT feedback can be modulated by the wind profiles that steer the convective structures and affect the organization and lifetime of PPT while its strength is likely controlled by the convective instability that leads to the generation of PPT. We tested the sensitivity of the PPT anomaly patterns to the strength of the low-level (10 m) wind when CI occurred. The low-level wind was chosen because it is related to convection over subtropical South America \citep{Salio2007_mcs}, advecting heat and moisture from the Amazon and the LPB southward, and is strongly correlated with vertical wind shear in the lower troposphere. This classification is based on the same thresholds of background wind as in Section \ref{wind_section}.

The spatial location of the largest PPT anomaly shows a strong sensitivity to the background wind speed (Fig. \ref{ppt_wind}). The lowest tercile is associated with enhanced rainfall close to the initiation point as the lack of background wind weakens the propagation of convective cells and lack of shear reduces their lifetime \citep{RKWtheory_1988}. Similar to the idealized modeling study of \cite{froidevaux_wind_feedback_2014}, the convective cycle during weak wind speed is anchored to the CI location likely leading to negative SM-PPT feedback. The statistically significant and positive spatial correlation (+0.59) between composite LSTA (antecedent to CI) and PPT anomalies (in the 24 hours following CI) within a 0.5$^{\circ}$ x 0.5$^{\circ}$ box centered at the CI location (Table \ref{lst_ppt_corr}) confirms this negative feedback. In contrast to \cite{froidevaux_wind_feedback_2014} who showed clear positive feedback during stronger wind speeds, the translation of PPT occurs perpendicular to the direction of the low-level wind (as opposed to parallel). The highest tercile shows the largest PPT anomalies between 150-250 kilometers away from the CI location in the negative crosswind direction (Fig. \ref{ppt_wind}c) as stronger low-level wind enhances the propagation of convective features and stronger vertical shear in the environment leads to enhanced mesoscale convective organization and increased spatiotemporal extent of rainfall. The lagged spatial correlation between composite LSTA and PPT anomalies is statistically significant and weakly negative (-0.20) which implies a weak positive SM-PPT feedback near the CI location. 

\begin{table}[!ht]
\caption{Spatial correlation between composite LSTA and PPT anomalies in 0.5$^{\circ}$ x 0.5$^{\circ}$ box centered at the CI location}
\label{lst_ppt_corr}
\centering
\begin{tabular}{@{} l *4c @{}}
\toprule
\multicolumn{1}{c}{\textbf{Sub-class}} & \textbf{Thresholds} & \textbf{Pearson-r} & \textbf{P-value} & \textbf{Number of Cases}  \\ 
\midrule
Low wind             & 0-1.5 m/s & +0.59    & 9.45e-43          & 1824                    \\ 
Moderate wind     & 1.5-2.5 m/s & -0.05      & 0.24         & 1675                     \\ 
High wind           & 2.5-10 m/s & -0.20     & 3.56e-05            & 1246                   \\
\midrule
All events             &  & +0.23       & 5.03e-07            & 4745                    \\ 
 \bottomrule
\end{tabular}
\end{table}

In summary, we show distinct peaks in PPT anomalies 150 km away from the CI location for the highest wind tercile and closer to the initiation location for the lower terciles (Fig. \ref{ppt_wind}d). Our results show strong sensitivity of the peak PPT location and the sign of the SM-PPT feedback to the strength of the low-level wind.

\section{Discussion} \label{discuss}

There is a large pool of numerical simulation studies showing heterogeneous SM/LSTA conditions triggering convection over spatially drier/warmer soils due to induced mesoscale circulations \citep{ChengCotton2004, Adler_inhomogeneities_2011, Garcia_InducedFlows_2011, KangBryan2011, Rieck_Clouds_2014, Rochetin_Breeze_2017, lee_effect_2019}. Previous observational evidence has found that SM heterogeneity of the order of 30-50 km exerts a strong influence on CI \citep{taylor_frequency_2011, taylor_detecting_2015, barton_observed_2021}. Our results agree with past studies and show that the length scale of the surface heterogeneity linked to CI over subtropical South America can vary between 30-100 km based on the background static and dynamic conditions of topography, wind, CAPE and CIN. This study is the first assessment of preferential convection over drier soils near strong dry-wet SM boundaries for the South American continent. 

This observational evidence of enhanced CI over drier/warmer soils is supported by three different satellite datasets of land surface characteristics and advocates for a realistic representation of mesoscale SM boundaries for an adequate prediction of afternoon convection. In forecasting mode, near realtime information of SM/LST conditions can potentially enhance skill in predicting the spatial location and timing of convective storms as demonstrated by \cite{Taylor_nowcasting_2022}. Simulating the lifecycle of convective storms without adequately capturing the role of SM heterogeneity will result in biases in weather prediction and climate simulations. 

The observed role of spatially drier soils over complex topography in initiating convection through induced upslope flow against the downslope background flow is a novel finding of this study. Our results in Figures \ref{topo_main}, \ref{topo_highclasses} and \ref{topo_elev} suggest that larger scale surface heterogeneity can induce upslope flow and CI over complex terrain. In past modeling studies \citep[e.g., ][]{Imamovic_orography_2017}, the spatial scale of the induced mesoscale circulation is constrained by the relatively small domain size. The importance of the larger scale SM gradients warrants closer examination.

The strength and significance of the negative LSTA gradients in this study are strongly sensitive to the background vegetation density (Fig. \ref{veg_main}). Dense vegetation can a) homogenize the spatial variability in surface sensible heat fluxes by accessing deeper soil moisture \citep{Ivanov_vegetation_2010}, and b) enhance spatial gradients in heat fluxes through the effect of roughness length on boundary layer winds. While the observed negative LSTA gradient is statistically significant at the 95\% level for all terciles of EVI, its strength reduces to less than half for EVI greater than 0.4. This suggests - a) the important role of roughness effects due to spatial heterogeneity in vegetation density in providing convective forcing, and b) a limitation of this current framework in quantifying the surface heterogeneity-related convective forcing in the presence of dense vegetation. While some past work has linked local circulations developing over anthropogenic land surface heterogeneities to the observations of cumulus development over the U.S Central Plains \citep{Weaver_atmospheric_2001} and the Amazonia \citep{Baidya_Amazonia_2002}, further investigation of the observed role of vegetation heterogeneity in initiating convection is needed to improve understanding. 

The prevalence of negative SM-PPT feedback has been observed by quantifying afternoon rainfall over spatially drier soils in multiple regions around the world \citep{taylor_afternoon_2012, guillod_reconciling_2015, welty_increased_2020}. When dry SM anomalies lead to enhanced PPT, the feedback is classified as negative. This is contrary to the positive SM-PPT feedback shown by a plethora of observational \citep{Findell_probability_2011, Koster_AGCM_2003} and global climate model studies with parametrized convection \citep{guo_glace_2006, taylor_afternoon_2012}. Our results show observational evidence that while convection can preferentially initiate over drier soils, the location of peak PPT can vary based on the background low-level wind. The description of SM-PPT feedbacks in a manner that captures the complete duration of convection from initiation to dissipation is lacking and needs to be addressed in future studies. 

The simple algorithm to detect CI in our study is likely missing the convective hotspot in the Sierras de Cordoba mountains located in northern Argentina. This region is known for the most intense convective storms on Earth \citep{Zipser_Intense_2006, Relampago_2021}. The fixed thresholds of cloud size, cloud core temperature and conservative filtering of cases with any cloud pixel in the vicinity fail to detect CI that occurs under conditions of strong wind shear and/or a pre-existing layer of cirrus clouds. Similar issues in designing a simple CI detection algorithm for satellite imagery have been discussed by \cite{Cancelada_BAB3T_2020, barton_observed_2021, Lee_convection_2021}. Future work can benefit from an adaptive algorithm that can capture the regional and latitudinal dependence of CI signature in satellite imagery.


\section{Conclusion} \label{conclude}

\begin{figure}[!h]
\centering
\vspace{\baselineskip}
\includegraphics[width=0.85\textwidth]{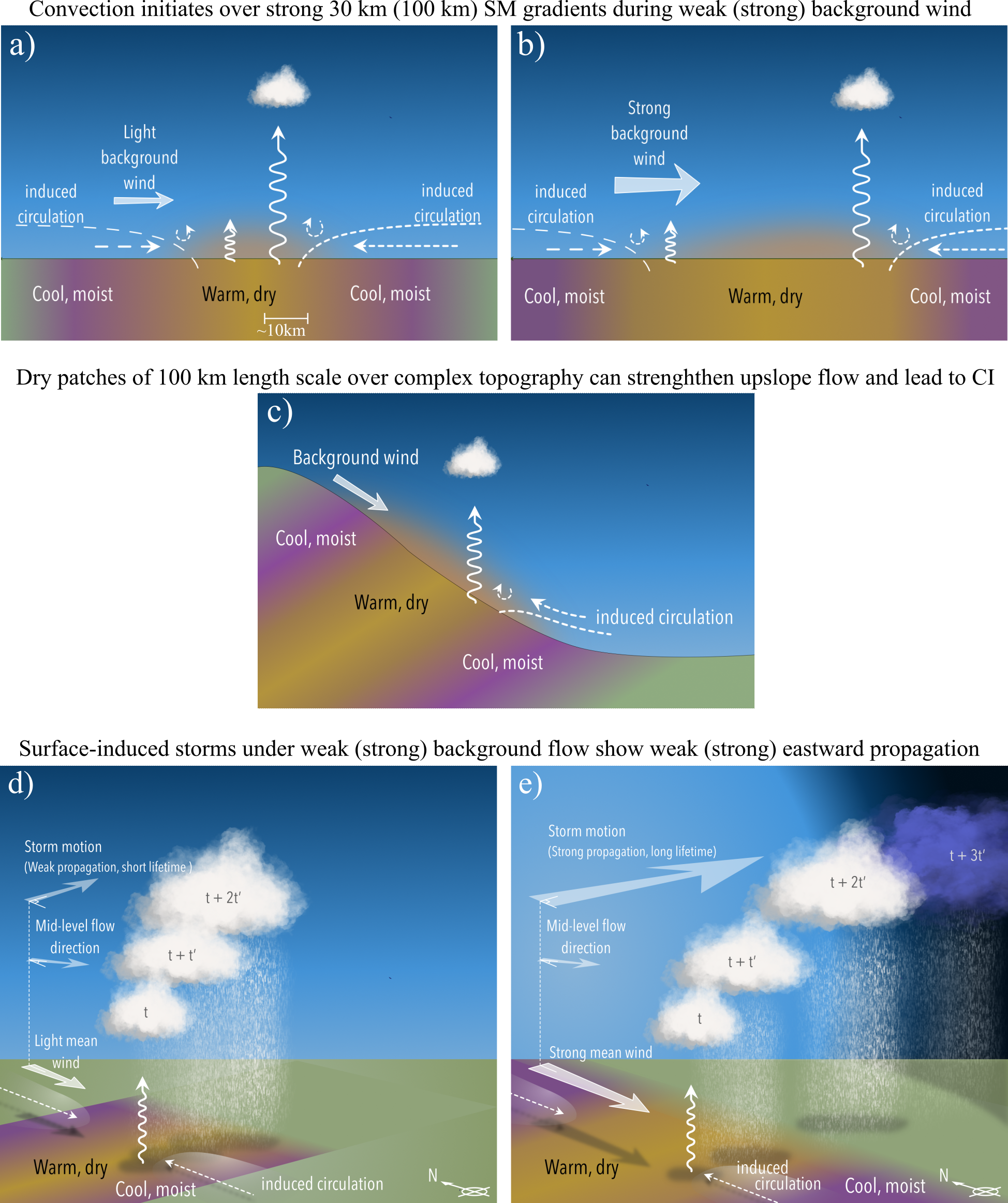}
\caption{\label{conceptual}a) Idealized SM heterogeneity-induced mesoscale transport (dashed arrows) under light synoptic wind (thick arrow) lead to ascent and convective initiation over the downwind edge of the dry patch, as hypothesized by \cite{taylor_frequency_2011}. b) Under strong background wind conditions, surface heterogeneity of a larger length scale ($\sim$100 km) can induce mesoscale flows strong enough to overcome the mixing effect of the stronger wind and lead to convection. c) Dry-wet soil patches located on elevated terrain directly heated by solar insolation can induce upslope flow to increase the convective forcing. d) Surface-induced storms that initiate under weak background flow result in peak rainfall accumulation over dry soils (negative SM-PPT feedback). e) Strong background wind leads to stronger propagation and longer lifecycle of convection resulting in PPT accumulation upto hundreds of kilometers east of the initiation location in subtropical South America.}
\end{figure}

The goal of this study is to observe and understand the role of surface heterogeneity at the mesoscale in initiating daytime convection under a range of environmental conditions and the consequent SM-PPT feedback during the warm season over subtropical South America. Major findings from this analysis are as follows:

\begin{enumerate}

\item{Convection over subtropical South America preferentially initiates on the dry side of strong dry-wet SM boundaries that correspond to positive alongwind SMA gradients and negative alongwind LSTA gradients on the order of tens of kilometers (Fig. \ref{conceptual}a).}

\item{The strength of the induced circulation is sensitive to the length scale of heterogeneity and the background wind. Convection is preferred over surface gradients of 30 km length scale during calm conditions with low-level wind speeds less than 2.5 ms$^{-1}$. Surface heterogeneity of a larger length scale (100 km) can lead to secondary circulations able to overcome the mixing effect of stronger background wind ($>$ 2.5 ms$^{-1}$) to initiate convection (Fig. \ref{conceptual}b).}

\item{In the presence of complex topography, SM patchiness of a larger length scale (100 km) can induce upslope flow (Fig. \ref{conceptual}c) that opposes the background downslope winds and leads to enhanced convection and rainfall over the eastern slope of the Andes High Plateau and the Brazilian Highlands.}

\item{The convective forcing due to surface heterogeneity is strongly sensitive to the presence of dense vegetation. While SM patchiness provides a strong and clear signal of convective forcing over sparse vegetation, roughness effects are likely crucial in the presence of dense vegetation.}

\item{Under convectively unfavorable conditions, strong gradients over a larger length scale can - a) induce mesoscale flows that concentrate CAPE within local regions and increase the likelihood of deep convection, and b) locally reduce CIN through enhanced surface heating.}

\end{enumerate}

The analysis of PPT anomalies over 24 hours following the tracked CI events shows:

\begin{enumerate}

\item{During weak background wind, the convective lifecycle is anchored to the initiation location and PPT occurs over the drier patch leading to negative SM-PPT feedback (Fig. \ref{conceptual}d).}

\item{Stronger background wind and shear support the propagation of convective structures hundreds of kilometers away from the initiation location, thus, confounding the sign of the SM-PPT feedback at convective scales (Fig. \ref{conceptual}e).}

\end{enumerate}

Our results show that mesoscale circulations induced by SM heterogeneity of the order of tens of kilometers represent an important forcing for convection over subtropical South America. The evaluation of SM-PPT feedbacks at convective scale for CI over drier soils is confounded due to the sensitivity of convective propagation to the background wind. This represents a major challenge as parametrized convection in climate models does not account for these subgrid circulations making it difficult to reproduce the observed diurnal cycle of precipitation. A better understanding of SM-PPT feedbacks is crucial for land management, adaptations to future climate change and our ability to forecast weather and make climate projections. Convection-permitting climate simulations that can explicitly capture the effect of surface heterogeneity in boundary layer transport and moist convection are necessary for accurately simulating the terrestrial water and energy cycles.

\clearpage
\acknowledgments
DC was funded by the Future Investigators in NASA Earth and Space Science and Technology (FINESST) Award 80NSSC19K1352. FD was supported by the National Science Foundation (NSF) Award 1852709. CMT and CK acknowledge funding from the UK Natural Environment Research Council (NERC) under the LMCS project (NE/W001888/1). SWN acknowledges funding by NASA Precipitation Measurement Missions Grant 80NSSC19K0713. The authors declare no conflict of interest. 

%
%
\datastatement
All datasets used in this study are publicly available. The information provided in Table \ref{datasets} can help readers locate and download the data used in this analysis: 

\begin{table}[!ht]
\caption{Additional information about the datasets used in this study.}
\label{datasets}
\resizebox{\textwidth}{!}{
\begin{tabular}{llll}
\hline
\textbf{Variable} & \textbf{Dataset name}                               & \textbf{Shortname}      & \textbf{Download Source}                                                                   \\ \hline
Infrared T$_{b}$  & NCEP/CPC L3 Merged IR V1                            & GPM\_MERGIR             & https://disc.gsfc.nasa.gov/datasets/GPM\_MERGIR\_1/summary                                 \\
Microwave SM      & SMAP Enhanced L3 Radiometer EASE-Grid SM V3         & SPL3SMP\_E              & https://nsidc.org/data/spl3smp\_e/versions/5                                               \\
Infrared LST      & MODIS/Terra LST L3 V061                             & MOD11C1                 & https://lpdaac.usgs.gov/products/mod11c1v061/                                              \\
Microwave LST     & AMSR2/GCOM-W1 downscaled LPRM L2B V001              & LPRM\_AMSR2\_DS\_SOILM2 & https://disc.gsfc.nasa.gov/datasets/LPRM\_AMSR2\_DS\_SOILM2\_001/summary                   \\
Wind, CAPE, CIN   & ERA5 hourly data on single levels                   & ERA5                    & https://cds.climate.copernicus.eu/cdsapp\#!/dataset/reanalysis-era5-single-levels?tab=form \\
Precipitation     & GPM IMERG Final Precipitation L3 V06                & GPM\_3IMERGHH           & https://disc.gsfc.nasa.gov/datasets/GPM\_3IMERGHH\_06/summary                              \\
Elevation         & Global Multi-resolution Terrain Elevation Data 2010 & GMTED2010               & https://www.usgs.gov/coastal-changes-and-impacts/gmted2010                                 \\
Lake mask         & Terra/MODIS land water mask V6                      & MOD44W                  & https://lpdaac.usgs.gov/products/mod44wv006/                                               \\
EVI               & MODIS/Terra Vegetation Indices                      & MOD13C2                 & https://lpdaac.usgs.gov/products/mod13c2v006/                                              \\ \hline
\end{tabular}}
\end{table}








%



\bibliographystyle{ametsocV6}
\bibliography{references}

\end{document}